\documentclass[fleqn,usenatbib]{mnras}

\usepackage{newtxtext,newtxmath}

\usepackage[T1]{fontenc}

\DeclareRobustCommand{\VAN}[3]{#2}
\let\VANthebibliography\thebibliography
\def\thebibliography{\DeclareRobustCommand{\VAN}[3]{##3}\VANthebibliography}

\usepackage{graphicx}	
\usepackage{amsmath}	
\usepackage{xspace}

\defcitealias{Murav}{M18}
\defcitealias{Muraveva_2025_metal}{M25}

\title[RR Lyrae Mid-IR PL independent of metallicity]{The independence of the mid-infrared RR Lyrae Period-Luminosity relation from metallicity from a study of three globular clusters in the Large Magellanic Cloud.}

\author[S. Ardern et al.]{
Steve Ardern,$^{1}$\thanks{E-mail: sa3129@bath.ac.uk}
V. Scowcroft,$^{1}$
S. Wuyts$^{1}$
\\
$^{1}$Department of Physics, University of Bath, Claverton Down, Bath, BA2 7AY, UK\\
}

\date{Accepted XXX. Received YYY; in original form ZZZ}

\pubyear{2026}

\begin{document}
\label{firstpage}
\pagerange{\pageref{firstpage}--\pageref{lastpage}}
\maketitle

\begin{abstract}
RR Lyrae are pulsating variable stars tracing old ($>10$~Gyr) stellar populations, and exhibit a strong mid-infrared Period-Luminosity (PL) relation that can be calibrated with known parallaxes to infer distances. Here we present Period-Luminosity relations and distance moduli for three isolated globular clusters in the Large Magellanic Cloud: Reticulum, NGC~1841, and NGC~1466. Our analysis uses legacy \textit{Spitzer Space Telescope} images obtained by the Carnegie RR Lyrae Program, and an internally self-consistent sample restricted to RRab stars, with cluster membership confirmed using Gaia DR3 proper motions and photometry. In the Spitzer 3.6~$\mu$m band, we simultaneously fit a PL for these clusters using a slope derived from Galactic Globular Clusters, yielding extinction-corrected distance moduli of {$18.47\pm0.09$} mag for Reticulum, {$18.29\pm0.09$} mag for NGC 1841, and {$18.61\pm0.09$} mag for NGC 1466. The latter two are the first RR Lyrae PL-based distance moduli for these clusters, and all three are consistent with literature values from other techniques. Additionally we fit a PL with slope as a free parameter, and find that this LMC-derived PL is consistent with that derived from Galactic GCs. Simultaneously fitting a PLZ for the three clusters yields a metallicity coefficient $c= -0.03\pm0.05~\mathrm{mag}~\mathrm{dex}^{-1}$ which can be considered a negligible dependence of the PL on metallicity ($|c| <0.1$ mag dex$^{-1}$). Modelling an intrinsic width, $W$, to the PL/ PLZ yields a consistent value $W\approx0.1\pm0.02~\mathrm{mag}$, suggesting intrinsic width is not driven by metallicity.
\end{abstract}




\section{Introduction}
\label{sec:introduction}

\subsection{RR Lyrae variables}
\label{sec:rrl_vars}

\begin{figure}
\centering
\includegraphics[width=1.\columnwidth]{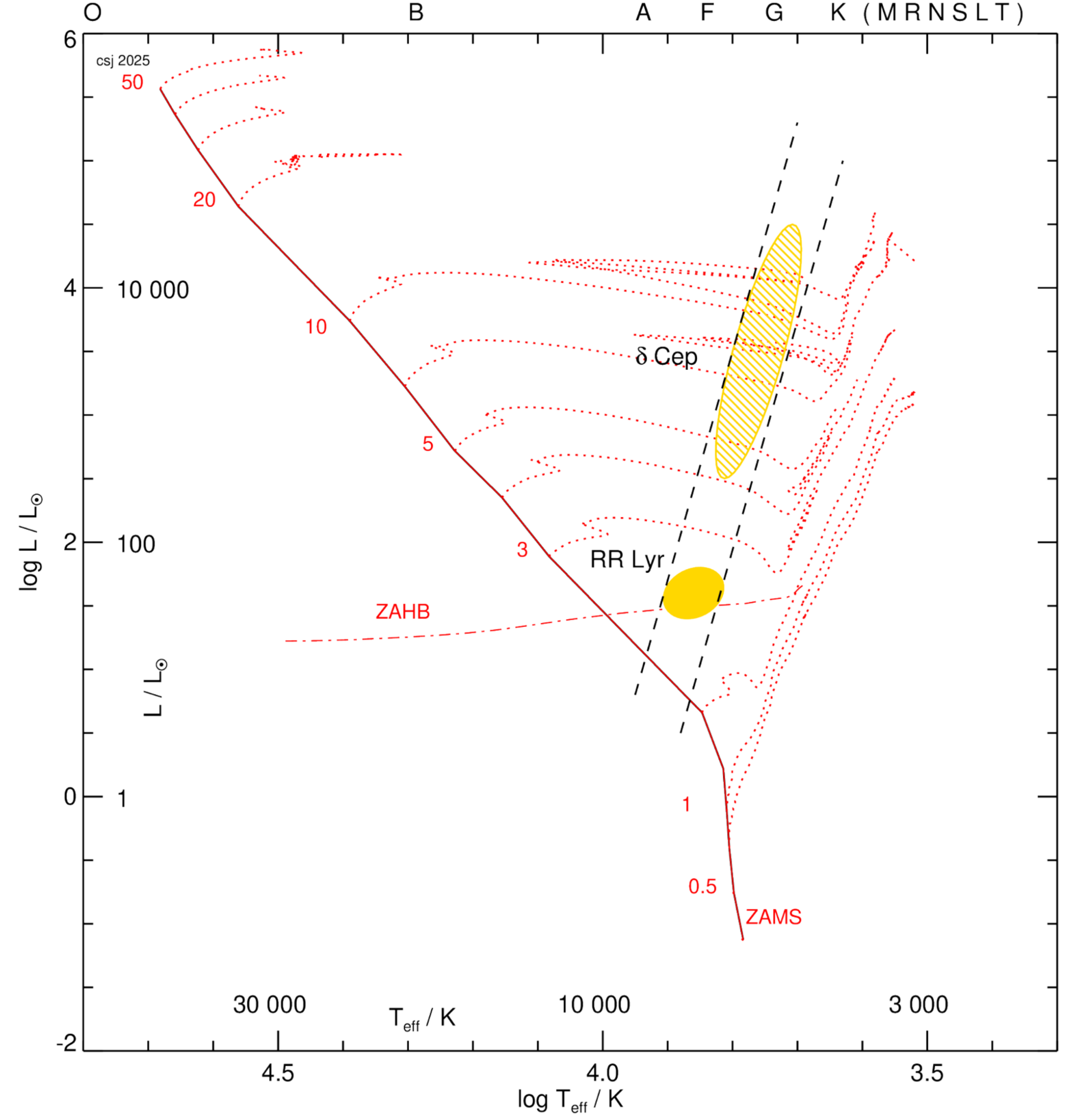}
\caption{Pulsation Hertzsprung-Russell Diagram \citep[adapted from][]{2016MNRAS.458.1352J} with permission. The solid red line shows the Zero Age Main Sequence (ZAMS), while selected evolutionary tracks are shown as dotted lines, with red numbers showing masses (M$_{\odot}$) for those tracks. The classical instability strip is shown by the dashed black lines. RRL are shown by a yellow-filled region, and are found where the Horizontal Branch crosses the instability strip -- the Zero Age Horizontal Branch (ZAHB) is shown as a red dot-dashed line. Cepheids are included as the yellow diagonally-hashed region for comparison.}
\label{fig:HR_Diagram}
\end{figure}

RR Lyrae (RRL) are pulsating variable stars with pulsation periods in the range 0.2 -- 1.0 days \citep{2009Ap&SS.320..261C}. They are evolved, low mass ($\thicksim$0.7 M$_{\odot}$) stars that have departed from the Main Sequence (MS), made their way up the Red Giant Branch (RGB), have undergone the He Flash and started burning He in their cores on the Horizontal Branch \citep[HB,][]{Caputo85}. They can be found in a narrow strip on the Hertzsprung-Russell (HR) diagram where the HB crosses the instability strip \citep{2009Ap&SS.320..261C}, as shown in Figure~\ref{fig:HR_Diagram}. 
Because of their low mass, RRL progenitors reside for at least 10 Gyr on the main sequence, making this the lower age limit for the onset of pulsational behaviour. The RRL phase takes place during a brief period in the long life of these stars; the maximum He burning time for a lower-mass HB star is around 100 Myr. However, the journey across the instability strip takes only part of this He burning lifetime, with the variable behaviour of these stars  lasting tens of Myr. For an individual star, the time during which variable behaviour is observed is dependent on where the star starts out on the HB, and whether this point is within the instability strip, or some location beyond the blue edge \citep{2007A&A...476..307L}. Having formed from molecular clouds that collapsed over 10 Gyr ago, RR Lyrae can be found in older regions of their host galaxies, including in globular clusters and galaxy haloes. These clouds collapsed to form stars early in the enrichment history of their host galaxies, meaning that RR Lyrae in globular clusters are generally low in metals. The old globular clusters of the LMC that are known hosts of RRL have mean [Fe/H] in the range -2.2 to -1.5 dex \citep{14clusters} -- Metal Poor to Very Metal Poor on the \citet{2005ARA&A..43..531B} scale.

\subsection{Horizontal Branch morphology and RRL}
\label{sec:HB_morph}
Observationally, Globular Clusters (GCs) that are more metal-poor host bluer HB stars, while more metal-rich clusters host redder HB stars. From a theoretical perspective, when initial chemical composition is fixed, the key driver for the colour difference becomes envelope mass, with thinner envelope stars being bluer, and thicker envelope stars being redder \citep{2019A&A...629A..53T}. There is, however, many decades of evidence to show that GCs with similar metallicities can exhibit different HB morphologies -- what has become known as the second parameter problem \citep{1967ApJ...150..469S}. Numerous theories have been proposed on the second parameter, including initial He fraction, dynamical effects related to cluster mass, dynamical evolution, or indeed some combination of factors \citep[see][for a review]{2009Ap&SS.320..261C}. However, recent observations point to mass loss on the RGB decreasing with metallicity \citep{2025ApJ...988..179L}, hence it is possible that RGB mass-loss could also be a fundamental factor in shaping a cluster's HB. The specific luminosity of the Horizontal Branch, with its narrow luminosity range in the \emph{V}-band, is directly related to the electron degeneracy of the He core of these stars prior to He-flash, which is itself a consequence of the rates of Main Sequence nuclear reactions \citep{2013osp..book.....C}. As the HB does not appear horizontal on a mid-IR colour magnitude diagram, the position of RRL on the HB is affected by not only $T_{\textrm{eff}}$, but also its mean mid-IR magnitude, and therefore pulsation period.

Like other HB stars, RRL evolve and migrate from the zero-age horizontal branch (ZAHB), undergoing changes in $T_{\textrm{eff}}$ and $\log(g)$ which will impact both period $(P)$ and amplitude $(A)$ of pulsation. However, empirical evidence is lacking on how HB evolution impacts the position of an RRL on the Bailey diagram \citep{2020ApJ...896L..15B}.

\subsection{RR Lyrae as standard candles in the infrared}
\label{sec:rrl_mir_candles}
In the \textit{V}-band, RR Lyrae variables do not exhibit a Period-Luminosity relationship (PL); instead there is a \textit{V}-band absolute magnitude-metallicity relation \citep{2003MNRAS.344.1097B}. However, \cite{1986MNRAS.220..279L} identified that there is a near-infrared (near-IR) PL for RRL. 

Theoretical studies in the following decades \citep[\textrm{e.g.}][]{2001MNRAS.326.1183B, 2004ApJS..154..633C, 2015ApJ...808...50M} provided a framework for multi-band Period–Luminosity–Metallicity (PLZ) relations, predicting steeper slopes and tighter correlations toward longer wavelengths. These predictions were confirmed observationally in the mid-infrared by Madore et al. (2013). The mid-IR PL has subsequently been used as a distance measure, \citep[e.g.,][ hereafter M18]{Murav}. Obtaining robust PL relations and hence distances for RRL stars presents a number of key opportunities: in this paper we discuss how RRL can be used in a Cepheid-independent distance scale and investigate the role played by metallicity in their PL.

\subsubsection{Cepheid-independent distance scale}
\label{sec:no_cep_ladder}
Most cosmic distance ladder measurements of the Hubble constant, $H_0$, are tied to Cepheids in the Milky Way and Magellanic clouds \citep[\textrm{e.g.}][]{2024IAUS..376....1F, Riess2022H0, 2025ApJ...992L..34R}. Geometric distances to local Cepheids are used to calibrate the Leavitt law (Period-Luminosity relation), enabling calibration of Type Ia Supernovae (SN Ia) absolute magnitudes in galaxies which host both Cepheids and SN Ia. This calibration can then be applied to progressively further galaxies, leading to a final $H_0$ measurement from galaxies well into the Hubble flow.

However, there is ongoing tension between the local universe measurements of $H_{0}$ and early universe Cosmic Microwave Background (CMB) measurements. Results from \emph{Planck} suggest $H_0 = 67.4 \pm 0.5$~km~s$^{-1}$~Mpc$^{-1}$ \citep{PlanckH0}, while \citet{Riess2022H0} find a significantly higher value of $H_0 = 73.04 \pm 1.04$~km~s$^{-1}$~Mpc$^{-1}$ using their Cepheid-based distance ladder. This discrepancy, dubbed the ``Hubble tension'' is seen (to varying degrees of significance) across multiple experiments, and the reader is directed to \citet{Verde2024} for a comprehensive review. 

One possible source of the Hubble tension lies in Cepheid-specific systematics that could be avoided with a Cepheid-independent distance ladder. Cepheids are young stars, hence the Cepheid distance scale is limited to observations of galaxies with recent star formation -- there is no such requirement for RRL host galaxies to have had recent star formation, due to the age of these stars. Moreover, Cepheids are mostly found in the discs of galaxies, and populations of Cepheids are therefore best observed in galaxies that present face on to us -- with RRL being found in galactic haloes their host galaxies can be in any orientation to be observationally useful. Hence an RRL distance scale can use all types of galaxy, in all orientations, dramatically improving the number of useful galaxies in a given volume, when compared to the Cepheid distance scale. The low-density environments in which RRL can be found also means that interstellar extinction is minimised. In this work, we analyse RR Lyrae in three old globular clusters in the Large Magellanic Cloud (LMC) using legacy \textit{Spitzer Space Telescope} images obtained as part of the Carnegie RR Lyrae Program (CRRP, PI Freedman, Program ID 90088). The CRRP was designed to develop the first rung of a Population II distance ladder to establish a Cepheid-independent distance scale \citep{2016ApJ...832..210B}.

\subsubsection{Role of metallicity in infrared Period-Luminosity relations}
\label{sec:Role of met}
There has been much debate in the literature on the significance of the metallicity coefficient ($c$) in the Period-Luminosity-Metallicity relation (PLZ) of the form\begin{equation}
    \label{eq:Influence of gamma}
    M=a \log{P} +b +c\mathrm{[Fe/H]} \mathrm{,}
    \end{equation}where $M$ is the absolute magnitude, $a$ the slope, $P$ the period of pulsation in days, $b$ the zero point, $c$ the metallicity coefficient, and [Fe/H] represents the metallicity in dex. Theoretical work suggests a metallicity term in the zero point of PL relations of RRL in the near-IR \citep[e.g.][]{2001MNRAS.326.1183B, 2004ApJS..154..633C,2015ApJ...808...50M}. \citet{Neeley} suggests that ignoring the RRL metallicity spread of $-2.6<$ [Fe/H] $<0.1$ dex yields an intrinsic width of 0.13 mag in the PL, while accounting for it results in a scatter of just 0.02 mag. Some empirical studies agree with the theory 
\citep[e.g.][]{2014MNRAS.439.3765D, Neeley2017zero_point, 2023ApJ...944L..51B} finding $c >0.1$ mag dex$^{-1}$. Other empirical studies on the other hand disagree with theory \citep[e.g][]{2008MNRAS.384.1583S, 2009A&A...502..505B, 2015ApJ...807..127M}, finding negligible dependence of the PL on metallicity ($|c| <0.1$ mag dex$^{-1}$). If there is indeed a dependence of the PL on metallicity, then accurate metallicities are required to make the RRL-based distance scale competitive. If metallicity is not significant, then the RRL-based distance scale becomes simpler and less costly to deploy. 

A non-negligible, positive value of $c$, would also result in decreased distance estimates compared to when metallicity is not considered. With a $c$ of +0.1 mag dex$^{-1}$, and a typical cluster metallicity of -1.5 dex, we would obtain distance moduli 0.15 mag closer. It is therefore essential we assess $c$ in order to obtain accurate and precise RRL distances.

\subsection{Paper overview}
In Section~\ref{sec:Data} we discuss the \textit{Spitzer} observations and \textit{Gaia} variable stars data used in this work, and our tests of cluster membership. Section~\ref{sec:Photom} describes our photometry methodology and how we obtain light curves and mean magnitudes for the RRab stars. In Section~\ref{sec:PL Relations} we report our measured PL relations and distance moduli for each cluster, comparing our results to the literature values. In Section~\ref{sec:PLZ relations} we show that the RRab PL can be considered independent of metallicity. In Section~\Ref{sec:IW} we  investigate whether the RRab PL is best described by including an intrinsic width term. Our conclusions are presented in Section~\ref{sec:Conc}.

\section{Archival Data}
\label{sec:Data}
\subsection{Warm \textit{Spitzer} observations}
\label{sec:maths} 
\begin{figure}
\centering
\includegraphics[width=\columnwidth]{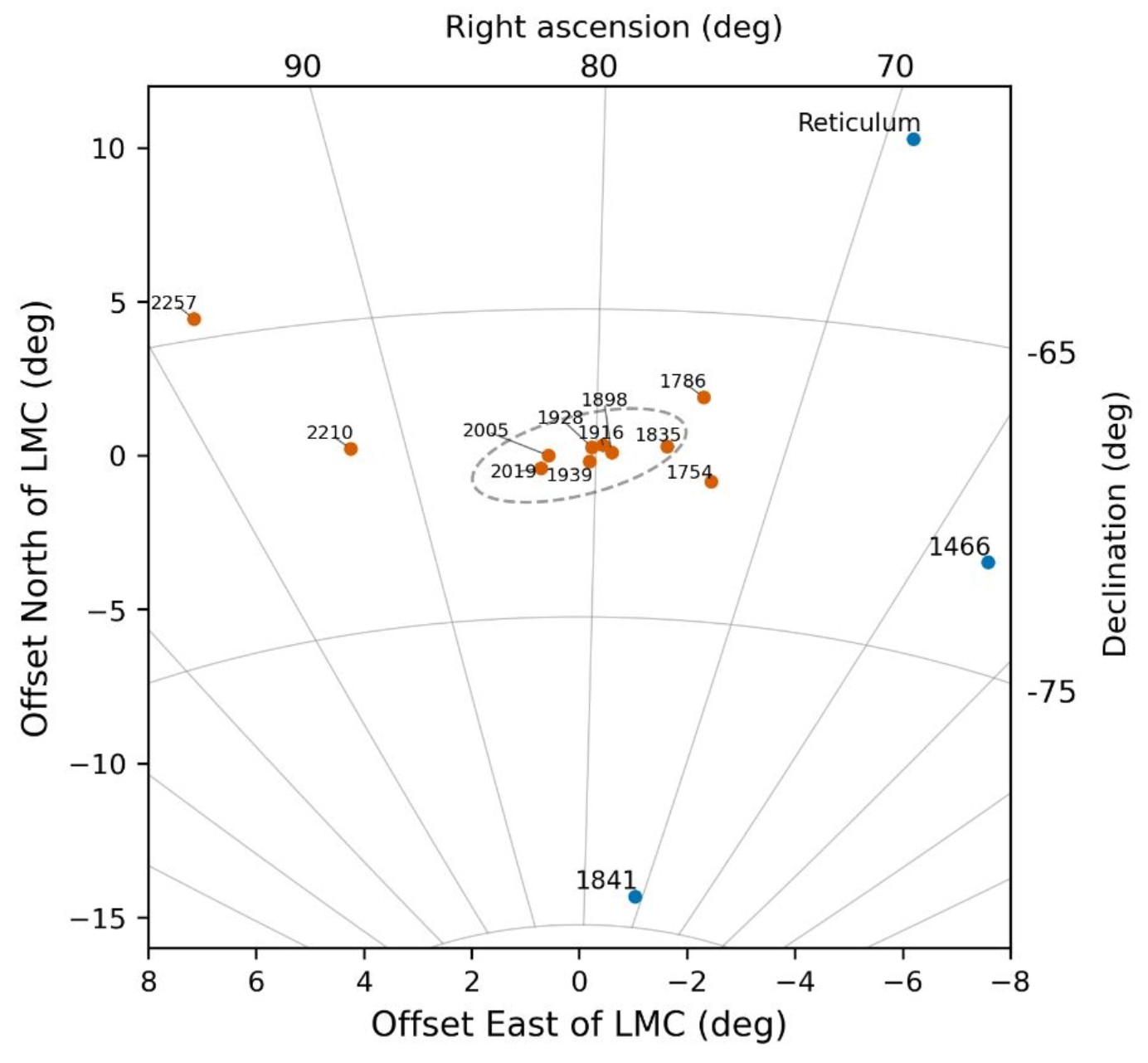}
\caption{Sky positions of the 14 old LMC clusters that are identified by \citet{14clusters} as known hosts of RR Lyrae. Numbers indicate NGC object identifiers. The three clusters used in this work are shown in teal, and the remaining clusters are shown in orange. The position of the central bar of the LMC is shown by the dashed oval \citep{2025ApJ...978...55R}. The three clusters used in this work are in the LMC's halo which has a low LMC field star density. The majority of the remaining clusters are located in, or close to, the central bar with its significantly higher LMC field-star density.}
\label{fig:Sky_positions}
\end{figure}

There are 15 LMC clusters old enough ($>10$ Gyr) to host RR Lyrae, and of these, 14 are known RRL hosts \citep{14clusters} as shown in Figure~\ref{fig:Sky_positions}. 
The Carnegie RR Lyrae Program conducted an observational campaign covering seven of these old globular clusters in the LMC. CRRP obtained images with the \textit{Spitzer Space Telescope}'s IRAC Instrument during the Warm \textit{Spitzer} Mission in both [3.6] and [4.5] bands. The work presented here uses [3.6] images of the three least-crowded CRRP clusters: Reticulum, NGC~1841 and NGC~1466. 

Each cluster was observed at 12 epochs across a 14~hr window, timed to optimize RR Lyrae phase coverage and to avoid aliasing. Observations for the clusters used in this work took place in November and December 2012 -- dates are shown in Table~\ref{tab:obs_dates}. Corrected basic calibrated data (CBCD) images were downloaded from the \textit{Spitzer} Heritage Archive and combined into single-epoch mosaics using MOPEX \citep{mopex} with a scale of 0.6 arcsec pixel$^{-1}$. In addition to the single-epoch mosaics, mosaicked location-correction and coverage images were created for each epoch. Mean exposure times for the resulting 7-dither mosaics are also shown in Table~\ref{tab:obs_dates}.

\begin{table}
    \caption{Observation dates \& mosaic exposure times per epoch for the three clusters.} 
    \label{tab:obs_dates}
    \centering
    \begin{tabular}{|c|c|r|}
    \hline
    {Cluster} & {Observation Date} & {Mean Mosaic}\\
    {} & {} & {Exposure Time (s)}\\
    \hline
    Reticulum & 2012 Nov 27 & 169.40\\
    NGC~1841 & 2012 Nov 27 & 23.49\\
    NGC~1466 & 2012 Dec 30 & 23.59\\
    \hline    
    \end{tabular}
\end{table}

\subsection{RR Lyrae catalogue}
\subsubsection{Restriction to RRab stars}
RRL can be split into distinct observational types according to their pulsation mode: fundamental mode pulsators (RRab), first-overtone pulsators (RRc), and those that pulsate simultaneously in the fundamental and first-overtone modes (RRd) \citep{Catelan_Smith}. A distinction can readily be made between RRab and RRc using a Bailey Diagram which plots \emph{V}-band amplitude against pulsation period, with RRc being grouped closer to the origin than RRab \citep{2011rrls.conf...17S}, while RRd can be identified via Fourier analysis of their light curves \citep[\textrm{e.g.}][]{2021MNRAS.507..781N}. While many previous works have considered mixed samples of RRab/RRc stars, here we limit our analysis to RRab alone.

RRab, RRc and RRd pulsators obey PL relations in the infrared. RRc and RRd overtone pulsators can be included on the same PL relationship as RRab stars in order to increase sample size, only when $\log P$ of these stars is shifted by a correction factor -- a process known as \textquoteleft Fundamentalization\textquoteright~\citep{Iben_1974}. However, the fundamentalization approach has recently been challenged. \cite{Narloch_fundametal} identified that fundamentalization does not correctly align RRab and RRc pulsators. Both \cite{Zgirski_2023} and \citeauthor{Narloch_fundametal} identified that slopes for RRab and RRc PL relations are different, and that this introduces systematic errors, with overestimation of RRab luminosities. \cite{Zgirski_2023} and \cite{Nemec_2024} suggest modifications to the fundamentalization process, but neither of these approaches removes the systematic resulting from the difference in temperature of the different classes of RR Lyrae being incorporated into the PL. This work focuses exclusively on RRab stars to avoid these systematics. Restricting our selection to RRab stars does not have a significant impact on the total number of targets as can be seen from Table \ref{tab:RRL_types} with only NGC~1466 hosting RRd, while Reticulum and NGC~1841 host only five and three RRc respectively within the three arcmin field of view. Only NGC~1466 hosts a significant number of RRc.

\begin{table}
	\centering
	\caption{RR Lyrae pulsational types in the three clusters from the {\tt{gaiadr3.vari-rrlyrae}} table within 3 arcmin of the cluster centres}
	\label{tab:RRL_types}
    {\begin{tabular}{lcccr} 
		\hline
		Cluster& RRab & RRc & RRd & Total RRL \\
        \hline
		Reticulum & 22 & 5 & 0 & 27 \\
        NGC~1841 & 14 & 3 & 0 & 17 \\ 
		NGC~1466 & 29 & 19 & 2 & 50 \\
		\hline
	\end{tabular}
    }
\end{table}

\subsubsection{\textit{Gaia} variable stars data}
\label{sec:GVS}
For consistency across the clusters, the catalogue RRab stars used in this work are drawn from the \textit{Gaia} DR3 archive \citep{2016A&A...595A...1G, 2023A&A...674A...1G}. For each cluster, all \textit{Gaia} DR3 sources within 6 arcmin of the centre of each cluster were selected. Additionally, any of these objects designated as RRab in the \textit{Gaia} DR3 {\tt{gaiadr3.vari-rrlyrae}} table within 3 arcmin of the cluster centres had further variable star parameters downloaded (including period and Fourier paramters of fitted light curves). The 3 arcmin criterion was a practical consideration relating to the area of the high SNR region in the \textit{Spitzer} mosaic images that would enable accurate photometry of these variables. The 6 arcmin radius was selected for all objects in order to provide a larger stellar sample to test against for cluster membership.

\subsubsection{Cluster membership confirmation tests}
\label{sec:Membership}
We conducted membership tests to confirm the \textit{Gaia} RRab used in this work are members of their respective clusters. Membership testing is a standard tool in the armoury of galactic archaeologists to group stars with similar origins. A number of different tests can be run, but perhaps the most straightforward method, and one which has been used for many decades, is to compare the proper motion of stars in RA and Dec using a Vector Point Diagram \citep{1958AJ.....63..387V}. This approach has been facilitated in recent years by the high-precision astrometry from the \textit{Gaia} survey \citep[e.g.][]{2019MNRAS.488.3024B, 2025A&A...695A..88G}. 

\citet{2019MNRAS.488.3024B} tested a number of approaches to determine cluster membership from \textit{Gaia} data. Interestingly, they found that uncertainty in \textit{Gaia} parallaxes made parallax measurements non-viable as a clustering parameter. Their preferred method applied a clustering algorithm to pre-cleaned 4-D data: X and Y sky-positions in gnomonic projection (to preserve local angular relationships and thereby avoid distortions), and proper motions $\mu_{\alpha \star}$ (proper motion in RA corrected for Dec) and $\mu_{\delta}$ (proper motion in Dec). They then fit a 4-D Gaussian to the identified candidates and rejected candidates at the $3\sigma$ level. Finally they used the sky distribution, Vector Point Diagram, and Colour Magnitude Diagram to check by eye that retained candidates were consistent with cluster membership.

\begin{figure*}
\centering
\includegraphics[width=\textwidth]{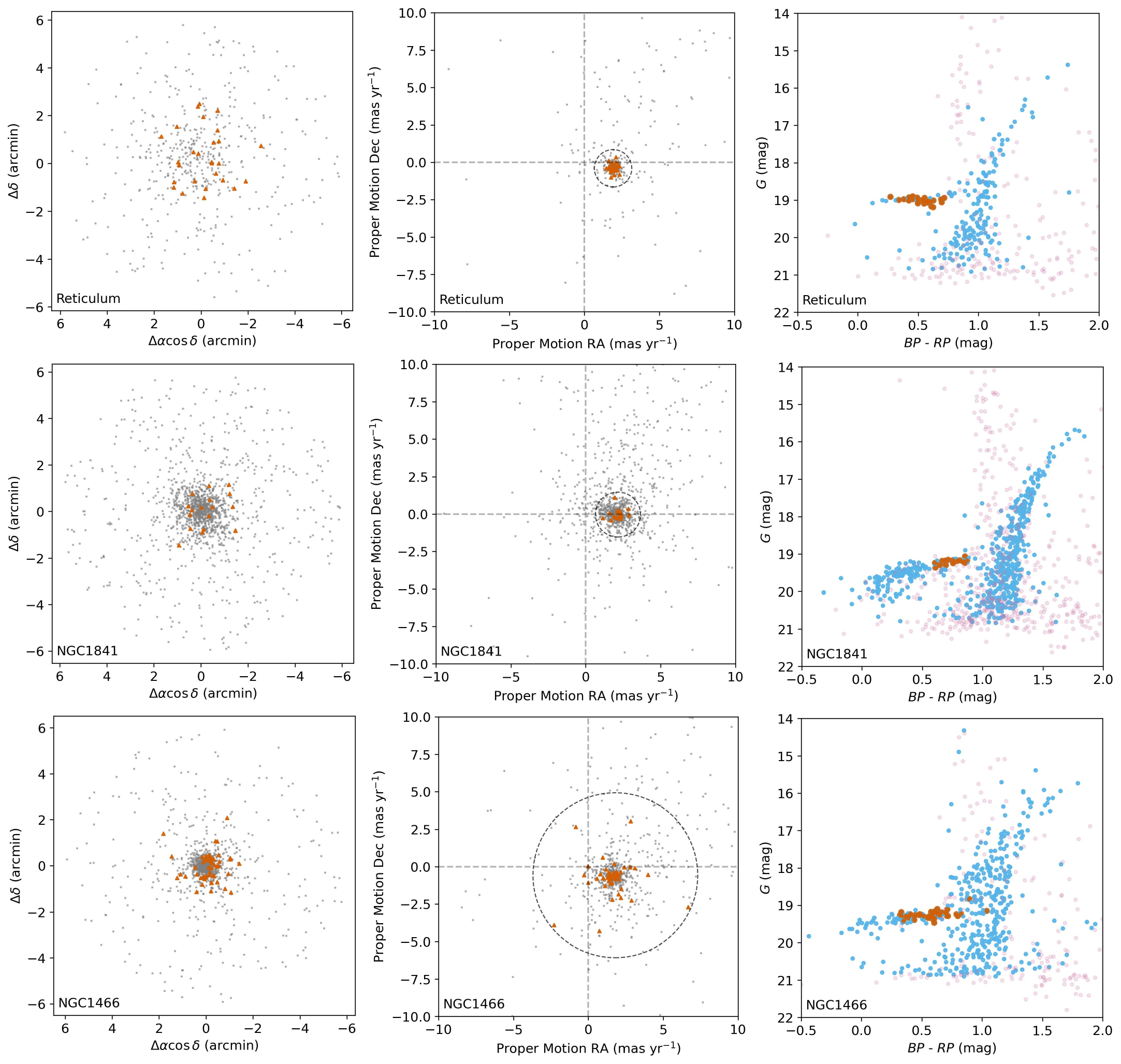}
\caption{Sky position, proper motion and colour-magnitude diagrams for the three clusters. Sky positions (left) are plotted in gnomonic projection ($\Delta \alpha\cos\delta~~\mathrm{versus}~~\Delta \delta$) to give undistorted representation of positions on the sky, grey dots represent all \textit{Gaia} objects to 6 arcmin, orange triangles represent \textit{Gaia} RRab to 3 arcmin of cluster centre. Proper motions (centre) are shown in mas yr$^{-1}$, grey dots represent all \textit{Gaia} objects to 6 arcmin, orange triangles represent \textit{Gaia} RRab to 3 arcmin of cluster centre. The black dashed circle shows the region of proper-motion space used for cut. Colour-magnitude diagrams (right) show RRab to 3 arcmin as orange points, \textit{Gaia} objects within 6 arcmin are represented by teal points where they were included as cluster members, and pink points where they were excluded by a cut in proper motion. For Reticulum and NGC~1841 the RRab can clearly all be seen to be members of their cluster's Horizontal Branch. For NGC~1466 there are two outliers at the red end (high \textit{BP}-\textit{RP}) of the Horizontal Branch; neither of these RRab were recovered by our photometry, hence all our recovered RRab for NGC~1466 are also members of the cluster's HB.}
\label{fig:Reticumlum_membership}
\end{figure*}

The clusters in our study are well isolated, and hence a simplified version of the \cite{2019MNRAS.488.3024B} methodology was used. For each cluster, we reviewed Vector Point diagrams and colour-magnitude diagrams of the \textit{Gaia} targets identified in Section~\Ref{sec:GVS}, making an initial cut in proper motion space, while also observing Horizontal Branch morphology. We then re-examined the cut versus uncut populations in proper motion space and on colour-magnitude diagrams to confirm our selection. This analysis is shown in Figure~\Ref{fig:Reticumlum_membership}, and confirms that, for Reticulum and NGC~1841, the \textit{Gaia} catalogue RRab stars within 3' of the cluster centre are found within the Horizontal Branch of the comoving population centred on the cluster, and should all be considered cluster members. For NGC~1466 there is uncertainty over cluster membership of two outliers at the red end (high \textit{BP}-\textit{RP}) of the Horizontal Branch; however neither of these RRab were recovered by our photometry, hence all our recovered RRab for NGC~1466 are also members of the cluster's HB. No cuts were made to the set of RRab stars as a result of these tests. 

\subsubsection{Sample cleaning}
\label{pulsation periods}
We did not use Lomb–Scargle periodograms to determine pulsation periods from the \emph{Spitzer} data itself because our 12 observational epochs span approximately a single pulsation cycle. In this regime, the period is only weakly constrained, as there is insufficient phase evolution to distinguish between nearby trial periods, leading to a strong covariance between the fitted period and phase. We therefore adopted the pulsation periods listed in the \textit{Gaia} DR3 {\tt{gaiadr3.vari-rrlyrae}} table, which are derived from the substantially more extensive Gaia time-series observations. These \textit{Gaia} periods are derived from the \textit{Gaia} light curves. Hence the quality of these light curves is central to our analysis. In the majority of cases, the \textit{Gaia} light curves are clean, constructed from \emph{ca.} 40 data points, and give unambiguous periods. We determined that a small number of \textit{Gaia} light curves are not clean, and these were cut from the study (two from Reticulum, two from NGC~1841, one from NGC~1466). The \textit{Gaia} data from \emph{ca.} 40 epochs spread over 34 months between 2014 July 25 and 2017 May 28, is also considerably closer in time to our \emph{Spitzer} observations in late 2012 than those from other possible sources of period data \citep[e.g.][]{1990AJ....100.1532W, 1992AJ....103.1166W, 1992AJ....104.1395W,  2004ApJ...610..269D}, hence we are confident that the periods obtained from \emph{Gaia} are the most appropriate for the remaining stars in our sample. The mean number of photometric observations, per RRab, used to construct the \emph{Gaia} light curves for each cluster are shown in Table~\ref{tab:Gaia_obs}. 
\begin{table}
	\centering
	\caption{Mean number of \emph{Gaia} light curve observations per RRab, per cluster}
	\label{tab:Gaia_obs}
    	\begin{tabular}{lr} 
		\hline
		Cluster& $\bar{n}$ observations \\
        \hline
		Reticulum & $39.4\pm1.5$ \\
        NGC~1841 & $43.7\pm1.5$ \\ 
		NGC~1466 & $42.6\pm1.9$ \\
		\hline
	\end{tabular}
    
\end{table}

\section{Photometry}
\label{sec:Photom}

\subsection{Use of \textsc{photutils}}
The first steps in producing a PL for any type of variable star are to select a stellar population, and to conduct multi-epoch photometry on those stars. A number of photometry codes have become recognised as standard practice including \textsc{daophot} \citep{Stetson}, \textsc{dolphot} \citep{2000PASP..112.1383D} and \cite{2006acs..rept....1A}'s code. Much of the functionality of these codes has recently been replicated within the \textsc{astropy} and \textsc{photutils} \textsc{python} packages, which have the merit of being developed within a 64-bit computing environment, appropriate for today's standard architectures. In this work, we build our photometric pipeline using \textsc{astropy} 5.1.1 and \textsc{photutils} 1.10. Both of these packages are updated regularly, and more recent versions are available, however the key components used in this work are the same. 

\subsection{Construction of EPSF model and EPSF photometry}
We conducted Effective Point Spread Function (EPSF) photometry, which compared to regular Point Spread Function (PSF) photometry, additionally seeks to account for pixel to pixel sensitivity variations, and variations in sensitivity across an individual pixel, with the aim of improved accuracy of both fluxes and centroids for undersampled images. The production of our EPSF model employed multiple steps to ensure high-precision. 

\subsubsection{Selection of best image}
We selected the best image from the 36 mosaic images (12 epochs, three clusters) to construct our EPSF model. The same 7-image dither pattern was used for all epochs, for all clusters, allowing a single mosaic to be used across all images. To confirm the validity of this approach we reviewed EPSFs prepared from different epochs for one cluster, and prepared from epochs in different clusters -- we found a mean per-pixel difference between EPSFs of $0.004\%$, and the resulting uncertainty in magnitudes was at the milli-mag level and can itself be considered negligible in comparison to dominant errors and is not included in our error analysis.

The exposure time for Reticulum was longer than for the other two clusters (see Table~\ref{tab:obs_dates}), and Reticulum is the least crowded; the epoch with the best SNR for Reticulum (Epoch 10) was therefore selected as the master epoch.

\subsubsection{Selection of EPSF model stars/ internal standard stars}
\label{sec:Selection process}
In this master epoch we selected bright, isolated, non-variable stars to use as our EPSF model stars, and to use as local standards for aperture photometry. Using the epoch's coverage file, the field was masked to the high SNR region, and the 250 brightest sources, meeting `sharp' and `round' limits, were identified by \textsc{photutils daostarfinder}. For each of the 250 sources, radial profiles of i) original image, ii) subtractions of the candidate local standards using a preliminary EPSF,  and iii) the subtraction of all stars using the preliminary EPSF were examined visually by two members of the team. This enabled deselection not only of candidates failing to meet the bright and isolated criteria, but also saturated sources, and those over or under-subtracted by EPSF photometry. The process was then repeated for the final selection of candidate internal standards. Stamps of a 25x25 pixel region around each source were then examined by eye to confirm no extended sources had been selected. \textsc{photutils} was used to compile a model of the EPSF using this set of EPSF model stars. The model fitting was bounded to positive flux values in the masked region. A 2-D plot of the EPSF model, and cross-sections of the model were inspected to check for anomalies.

2-D background subtraction was carried out on the mosaic to manage the residual cluster light at the centre of the images. This step was not required prior to creation of the EPSF model as the software extracts the EPSF stars from their local background. EPSF photometry was then conducted on all sources with a peak above 3$\sigma$ of the background.

Subtraction images were then made of: a) mosaic minus EPSF model stars, and b) mosaic minus all sources other than EPSF model stars. Radial profiles of the regions in these subtraction images centred on the coordinates of the EPSF model stars were visually inspected allowing further de-selection of those that were poorly subtracted, and those impacted by the subtraction of other stars, resulting in a revised set of EPSF model stars.

\subsubsection{Detailed review and revision of EPSF model}
\label{sec:EPSF model}
An EPSF model was built from the revised set of EPSF model stars in the master epoch using \textsc{photutils}. EPSF photometry was conducted as per Section \ref{sec:Calib to IRAC} for all epochs for Reticulum on all sources above 3$\sigma$ of the 2-D background, yielding instrumental magnitudes for these sources. Light curves were generated for the internal standard stars with the expectation that these would be flat across all 12 epochs. Where individual magnitudes in specific epochs were beyond 1$\sigma$ of the 12-epoch mean for an EPSF model star, that observation was flagged. A final master EPSF was then constructed omitting these specific flagged observations of EPSF model stars.

\subsubsection{EPSF photometry}
The selection process described in Section~\ref{sec:Selection process} was repeated for the other two clusters to create their own sets of internal standard stars. EPSF photometry was then conducted on all three clusters using these cluster-specific internal standard stars and the EPSF developed from the master epoch in Section~\ref{sec:EPSF model}.

\subsection{Calibration to \textsc{IRAC} photometric system}
\label{sec:Calib to IRAC}
\subsubsection{Aperture correction to the \textit{IRAC} photometric system}
Aperture photometry was carried out on the master epoch mosaic image using a [6, 6, 14] photometry system (6 pixel diameter aperture, with sky annulus from 6 to 14 pixels) -- this is equivalent to [3, 3, 7] in native pixels. Fluxes were converted to the [10, 12, 20] native pixel system using the aperture corrections provided in the \textit{IRAC} Instrument Handbook \footnote{IRAC Instrument Handbook. IRAC Instrument and Instrument Support Teams. Version 4.0 (Final), September 2021. \url{https://irsa.ipac.caltech.edu/data/SPITZER/docs/irac/iracinstrumenthandbook/}}. The [10, 12, 20] system is the one in which zero magnitude fluxes are defined by \citet{2005PASP..117..978R}. We also applied the CCD location correction provided in the \textit{IRAC} handbook to correct for non-uniform response across the images using the mosaicked location correction images. We found the weighted mean difference between our aperture photometry magnitudes in the \textit{IRAC} photometric system, and our EPSF photometry mags for the master epoch internal standards stars. This was used as our final systematic aperture correction to convert our EPSF magnitudes to the \textit{IRAC} photometric system for all objects in all epochs, for all three clusters.

\subsubsection{Pixel-phase correction}
\begin{figure}
\centering
\includegraphics[width=\columnwidth]{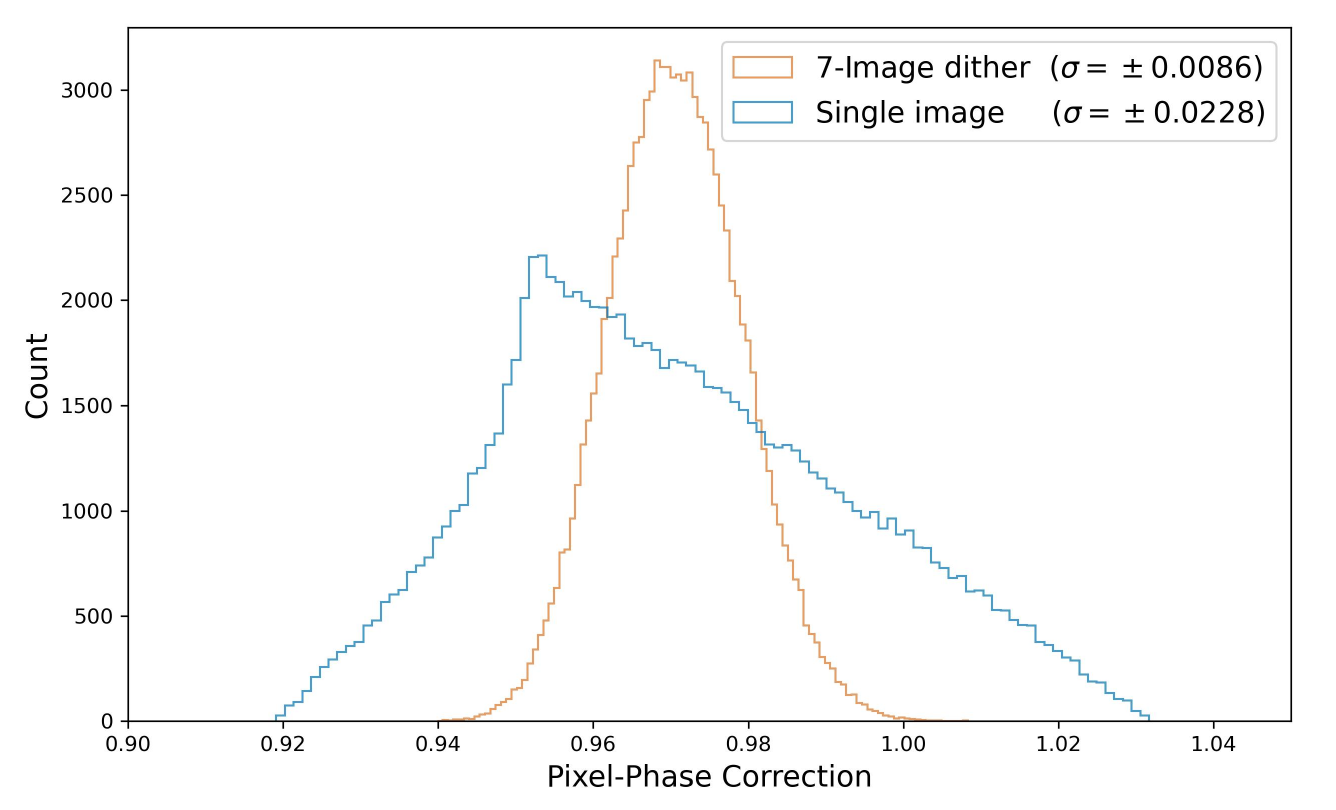}
\caption{Pixel-phase correction sampling for [3.6] photometry, showing histograms of 100,000 simulated single images and 100,000 normalised 7-image dithers. The individual images have a mean dimensionless pixel-phase photometric correction of $0.9707\pm0.0228$, while for the 7-image dithers the value is $0.9707\pm0.0086$.}
\label{fig:Pixel_phase}
\end{figure}

The precise location on a pixel where the PSF peaks, known as the pixel-phase, can have an impact on the measured flux density -- both the strong undersampling in IRAC, and variations in quantum efficiency across the pixel contribute to this. The \textit{IRAC} handbook states that intra-pixel variations on an individual [3.6] BCD image can be as much as 4 per cent peak to peak, however the average gain across a pixel is unity, hence pixel phase should also average out in a well-sampled mosaic. As shown in Figure~\ref{fig:Pixel_phase}, we confirmed this by randomly sampling the IRAC [3.6] intrapixel sensitivity map to compare the pixel-phase error expected from 100,000 single images, and 100,000 normalised 7-image dithers. We found the mean dimensionless pixel-phase correction for single CBCDs to be 0.9707 +/- 0.0228, and for the 7-image dither this was 0.9707 +/- 0.0086. The corrections are rolled-up in our aperture correction. The impact of the expected errors on these corrections in terms of magnitude errors is $\pm0.025$ mag for single images, for the 7-image dither this was $\pm0.009$ mag. This pixel-phase photometric error for the 7-image dither can be considered negligible in comparison to dominant errors and is not included in our error analysis. 

\subsection{Light curves and RRab mean magnitudes}
We used the \textsc{gloess} (Gaussian-windowed LOcal regrESSion method) to produce light curves for the recovered RRab stars. \textsc{gloess} was first used by \cite{2004AJ....128.2239P} and is described in detail in \cite{2017AJ....153...96M}. Magnitude and temporal data and \textit{Gaia} variable catalogue periods were fed to \textsc{gloess} which produced two and a half period phase-folded light curves, a selection of which are shown in Figure~\ref{fig:example_LCs}, with the full set provided in Appendix~\ref{APP_A}. \textsc{gloess} also provided flux-weighted mean magnitudes for each RRab; these are provided in Table~\ref{tab:mean_mags}. The principal sources of uncertainty in the flux-weighted mean magnitudes from this method are related to i) the amplitude of the light curve ($A$) and the number of observations used in the fit ($N_{\text{obs}}$) via:\begin{equation}
    \sigma_{i}=\frac{A}{\sqrt{N_{\text{obs}}}}
	\label{eq:mean_mag_err}
\end{equation} and ii) the uncertainty on the amplitude itself. Both these uncertainties were propagated through to the final uncertainty on RRab flux-weighted mean magnitudes.

\begin{figure*}
        \centering
        \includegraphics[width=\textwidth]{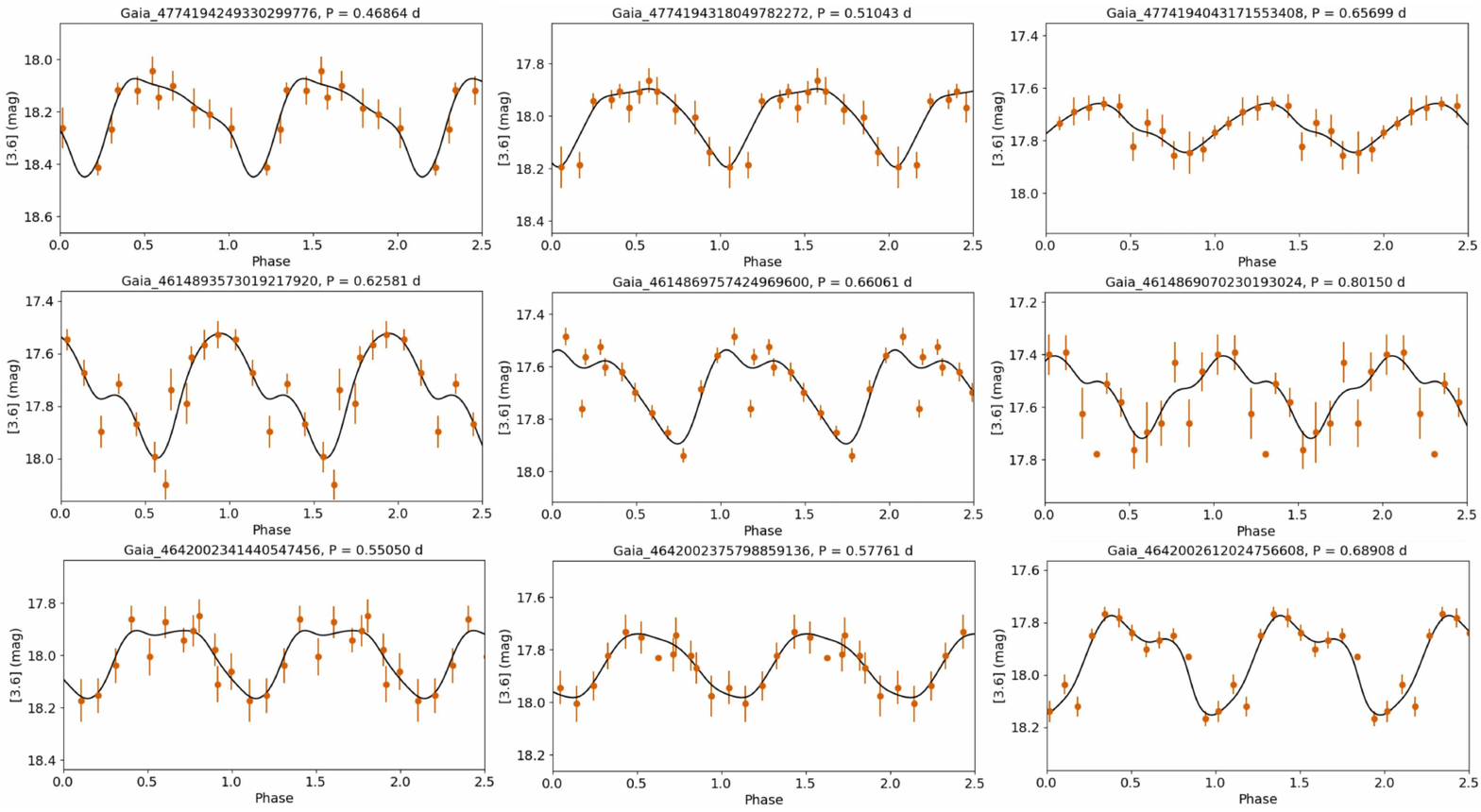}
        \caption{Example \textsc{gloess} fitted phase-folded light curves for RRab across a range of periods for the three clusters -- top row Reticulum, middle row NGC~1841, bottom row NGC~1466. The full set of recovered light curves is provided in Appendix~\ref{APP_A}. The orange data points and errorbars are our photometric results for each epoch, the black light curve is the fit provided by \textsc{GLOESS} to this dataset. A common magnitude range is used on all light curves. Note that some photometric data points do not have errorbars as the EPSF fitter was unable to provide them, however we have included these points for completeness.}
        \label{fig:example_LCs}
\end{figure*}

\section{Period-Luminosity relations}
\label{sec:PL Relations}
In this section we describe the different PL fitting approaches used in this work. We take the \cite{Neeley2017zero_point} [3.6] PL from the Galactic globular cluster M4 as fiducial as this is derived from RRab in M4, the closest Galactic globular cluster to our position, hence is expected to be more accurate than fitting a slope to our data from considerably more distant clusters in the LMC.

\subsection{Fixed slope PL fitting}
\label{sec:neeley17_slope}
We used a Bayesian linear regression model to simultaneously fit the basic PL relation \begin{equation}  \label{eq:BAsic PL equation}
m_{[3.6]} = a\log(P) + b
\end{equation}
to all three clusters. This was achieved by incorporating a parameter in the model to represent the offset for each cluster $(\Delta b)$, and fixing the offset for Reticulum to zero: 
\begin{equation}  \label{eq:Three cluster PL equation}
m_{[3.6]} = a\log(P) + (b + \Delta b_{\mathrm{cluster}})
\end{equation}

We applied flat priors of $\pm50$ on the intercept $b$, and $\pm10$ on $\Delta b$, to allow the full parameter space to be explored, and fixed the slope according to the RRab empirical PL relation derived from Galactic GCs in \cite{Neeley2017zero_point}: 
\begin{equation}   
\label{eq:Neeley equation}
M_{[3.6]} = -1.155(\pm0.089) -2.34(\pm0.14) \log(P).
\end{equation}

\begin{figure}
\centering
\includegraphics[width=\columnwidth]{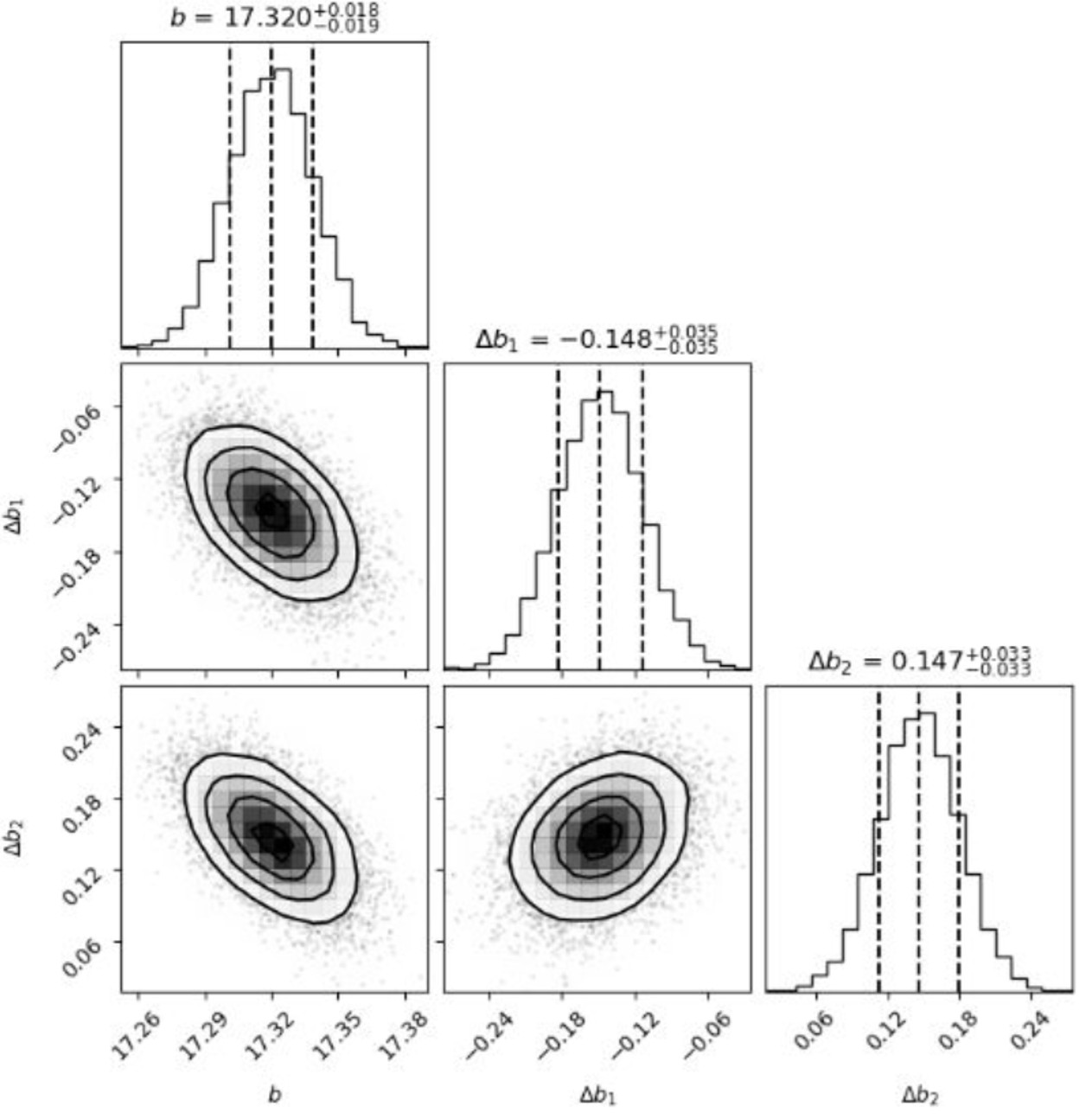}
\caption{Corner plot for the simultaneous three cluster fit PL with fixed slope showing representations of the posterior distributions of the parameters $b$ (intercept), $\Delta b_1$ (offset for NGC~1841), $\Delta b_2$ (offset for NGC~1466) and their cross-correlations.}
\label{fig:Three cluster PL}
\end{figure}

Markov Chain Monte Carlo (MCMC) sampling using the \textsc{emcee} \textsc{python} package \citep{2013PASP..125..306F} was used to derive the posterior distribution of the intercept. Our fit is weighted by the inverse variance of the data points. Brighter stars typically have lower photometric errors and can consequently be favoured in a weighted fit, leading to bright-biasing of the fit. However, RRab have a narrow band of intrinsic luminosity, and within each cluster our samples can be considered equidistant from us, hence bright-biasing should not be relevant in this case and a weighted fit can be employed. The fitting results are shown in Table~\ref{tab:reddening, extinction and distance} as the median of the posterior distribution with uncertainties reported as the 16th and 84th percentiles of the distribution. The corner plot for the fixed slope PL fit is shown in Figure~\ref{fig:Three cluster PL}.

\begin{table*}
	\centering
	\caption{Reddening, extinction and distance moduli.}
	\label{tab:reddening, extinction and distance}
	\begin{tabular}{lcccccccr} 
		\hline
		Cluster & $E(B-V)$ & $E(B-V)$~Ref. & $A_{[3.6]}$ & $b_{\mathrm{cluster}}$ (mag) & $ b_{\mathrm{cluster},0}$ (mag) & $\mu_{0}$ (mag) &  $n_{\text{RRab}}$ & $\Delta\mu$ (mag) \\
        &&&&$(b+\Delta b_{\mathrm{cluster}})$&&&&
        \\
        \hline
        Reticulum & $0.03\pm0.02$ & \citet{1992AJ....103.1166W} & $0.006\pm0.004$ & {$17.321\pm0.018$} & {$17.315\pm0.018$} & {$18.47\pm0.09$} & 18 & N/A \\
        NGC~1841   & $0.18\pm0.02$ & \citet{1990AJ....100.1532W} & $0.037\pm0.004$  & {$17.172\pm0.029$} & {$17.135\pm0.029$} & {$18.29\pm0.09$} & 9 & $+0.180$\\
        NGC~1466 & $0.09\pm0.02$ & \citet{1992AJ....104.1395W} & $0.018\pm0.004$ & {$17.468\pm0.026$} & {$17.450\pm0.026$} & {$18.61\pm0.09$} & 8 & $-0.135$  \\
        \end{tabular}
\end{table*}

\subsection{Extinction correction}
The apparent magnitudes we find for the RRab stars require correction for line-of-sight extinction to each cluster. We use the extinction laws of \cite{1989ApJ...345..245C} in the optical, and \cite{2005ApJ...619..931I} in the infrared to find
\begin{equation}   
\label{eq:Extinction}
A_{[3.6]} = 0.203  E(B-V),
\end{equation}
where $A_{[3.6]}$ is the [3.6] extinction and $E(B-V)$ is the colour excess. The self-consistent set of reddening values for the three clusters from \cite{1990AJ....100.1532W, 1992AJ....103.1166W, 1992AJ....104.1395W} were used to compute our extinction corrections and both are shown in Table~\ref{tab:reddening, extinction and distance}. We applied extinction corrections directly to the intercept $(b_{\mathrm{cluster}})$ to obtain the extinction corrected values $(b_{\mathrm{cluster},0})$ for each cluster shown in Table~\ref{tab:reddening, extinction and distance} using \begin{equation}
    b_{\mathrm{cluster},0} = b_{\mathrm{cluster}} - A_{[3.6],\mathrm{cluster}}
\end{equation}
Since the RRab in a given cluster have a common extinction, it acts as a shared systematic offset rather than star-to-star random noise. The uncertainty on the extinction is a correlated systematic uncertainty affecting the PL zero-point, rather than an independent uncertainty on each individual magnitude. We therefore fit the observed magnitudes directly in the PL and subsequently apply the extinction correction to the fitted intercept. This avoids unnecessarily inflating the individual magnitude uncertainties used for the PL fit, while still propagating the extinction uncertainty into the final distance modulus uncertainty.

\subsection{Zero point and distance modulus}
A Period-Luminosity relation must ultimately be anchored either to a geometric distance, or use another directly known distance to the objects in the study if it is to be used as an absolute distance measure. With LMC RRL \emph{Gaia} parallaxes being unsuitable to use for distances \citep{2021A&A...649A...7G}, the obvious candidate for this study is to again use the empirical PL relation provided by \cite{Neeley2017zero_point} for RRab in \textit{Spitzer} IRAC [3.6] shown in Equation~\ref{eq:Neeley equation}. The zero point for the \citeauthor{Neeley2017zero_point} PL was anchored to a sample of Galactic RR Lyrae with \textit{Gaia} DR2 distances. \citeauthor{Neeley2017zero_point} found good agreement between this zero point and the theoretical value they obtained from nonlinear hydrodynamical stellar modelling. 

\begin{figure*}
        \centering
        \includegraphics[width=\textwidth]{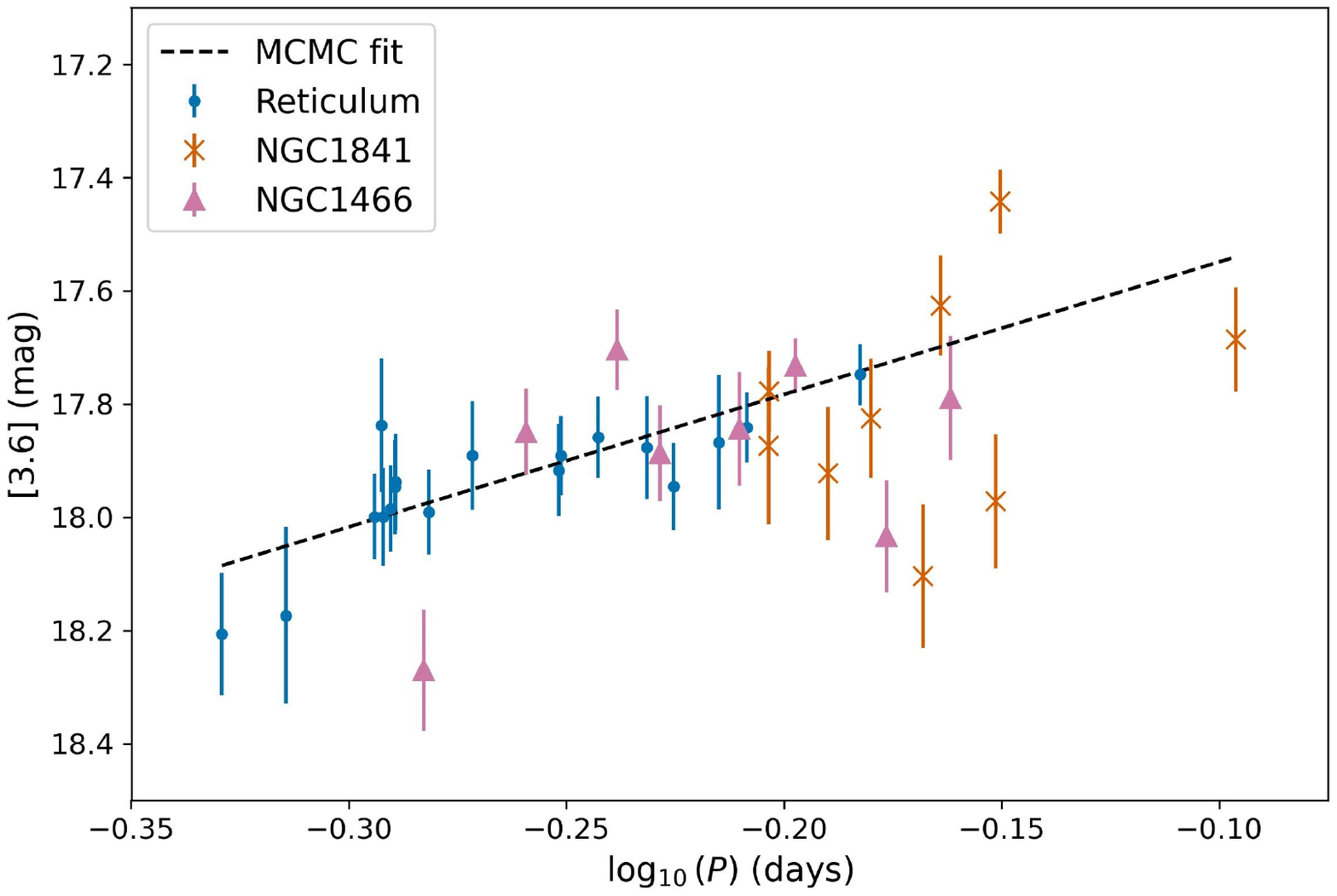}
        \caption{Period-Luminosity relation for RRab in the LMC globular clusters studied in this work. Apparent magnitudes were offset by the clusters' extinction-corrected distance moduli relative to Reticulum ($\Delta\mu$) to allow comparison as a single population: $\Delta\mu_{\text{NGC~1841}}=+0.180$ mag, and $\Delta\mu_{\text{NGC~1466}}=-0.135$ mag. It should be noted that there is a difference in the dynamic range of periods between the three clusters which may be reflective of different cluster ages and RGB mass-loss histories.}
        \label{fig:LMC_P-L}
\end{figure*}

The extinction-corrected intercepts for each cluster ($b_{\mathrm{cluster,0}}$) were converted to extinction corrected distance moduli ($\mu_{\mathrm{cluster,0}}$) using the \citeauthor{Neeley2017zero_point} zero point ($ZP_{\text{Neeley}}$) as: 
\begin{equation}   
    \label{eq:Mu calc equation}
    \mu_{\mathrm{cluster},0} = b_{\mathrm{cluster},0} - ZP_\mathrm{{Neeley}}
\end{equation}
The resulting distance moduli for the three clusters are shown in Table~\ref{tab:reddening, extinction and distance}. Differential extinction-corrected distance moduli relative to the Reticulum Cluster were calculated for NGC~1466 and NGC~1841 as:\begin{equation}      \label{eq:Diff dist moduli}
    \Delta\mu_\text{{cluster},0} = \mu_\text{{cluster},0} - \mu_\text{{Reticulum},0}
\end{equation}
The PL relations for the three clusters are shown in Figure~\ref{fig:LMC_P-L}, with NGC~1841 and NGC~1466 apparent magnitudes offset by their respective $\Delta\mu$ values to allow comparison as a single population. It is apparent from Figure~\ref{fig:LMC_P-L} that there is a difference in the dynamic range of periods between the three clusters which may be reflective of different cluster ages and RGB mass-loss histories.

\begin{figure*}
        \centering
        \includegraphics[width=1.\textwidth]{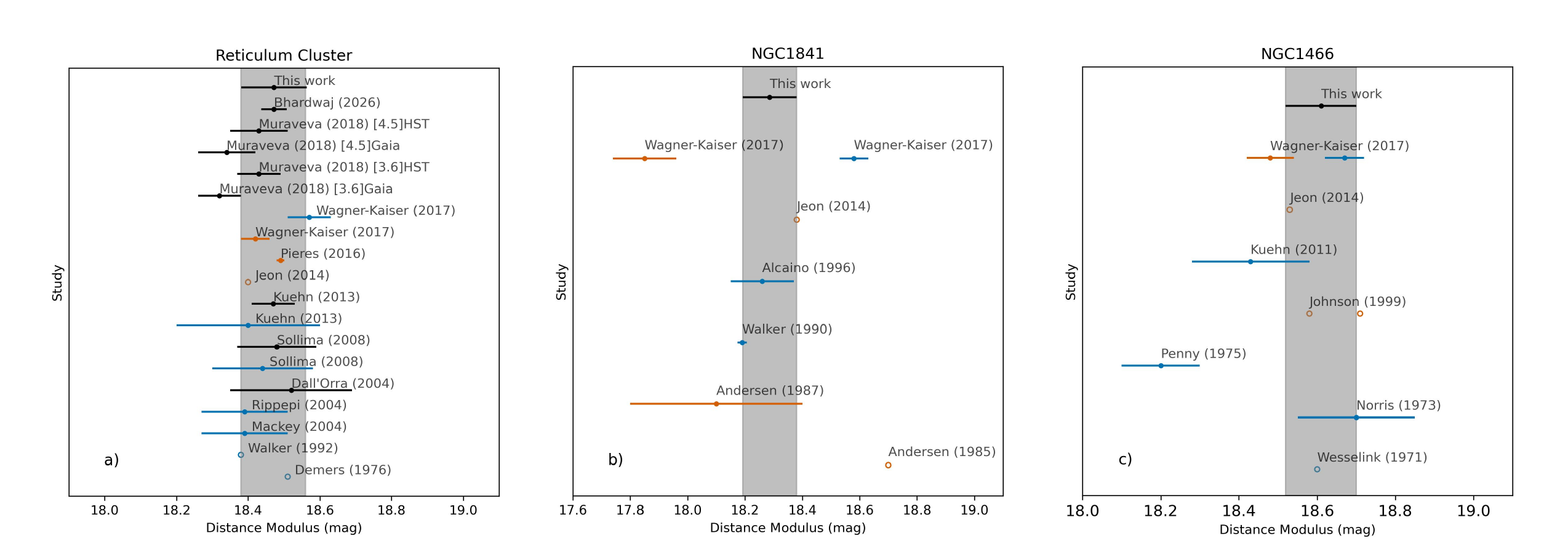}
        \caption{Comparison of distance moduli obtained by a range of methods for Reticulum, NGC~1841, and NGC~1466. Orange data points are from colour-magnitude diagram (CMD) fitting, black data points are from PL relations, and blue points are from Horizontal Branch/ RRL mags. Open circles denote data points with no uncertainty reported. Reticulum Cluster studies: \protect\cite{2026AJ....172..148B}, \protect\cite{Murav} (`\textit{HST}' and `\textit{Gaia}' labels denote how the zero-point was determined),  \protect\cite{2017MNRAS.471.3347W}, \protect\cite{2016MNRAS.461..519P}, \protect\cite{2014AJ....147..155J}, \protect\cite{2013AJ....145..160K}, \protect\cite{2008MNRAS.384.1583S}, \protect\cite{2004ApJ...610..269D}, \protect\cite{2004CoAst.145...24R}, \protect\cite{2004MNRAS.352..153M}, \protect\cite{1992AJ....103.1166W} and \protect\cite{1976ApJ...208..932D}. NGC~1841 studies: \protect\cite{2017MNRAS.471.3347W}, \protect\cite{2014AJ....147..155J}, \protect\cite{1996AJ....112.2020A}, \protect\cite{1990AJ....100.1532W}, \protect\cite{1987MNRAS.229....1A} and \protect\cite{1985A&A...150L..12A}. NGC~1466 studies: \protect\cite{2017MNRAS.471.3347W}, \protect\cite{2014AJ....147..155J}, \protect\cite{2011AJ....142..107K}, \protect\cite{1999ApJ...527..199J}, \protect\cite{1975MNRAS.172P..65P}, \protect\cite{1973ASSL...36..113N} and \protect\cite{1971MNRAS.152..159W}.}
        \label{fig:Retic_compare}
\end{figure*}

Figure~\ref{fig:Retic_compare} shows a comparison of the extinction corrected distance moduli for each cluster with those found by previous studies. The Reticulum cluster has been particularly well studied, and a comparison of the distance modulus obtained in this work with those found by others over the past 50 years provided in Figure~\ref{fig:Retic_compare}a, shows strong agreement between this work and other methods. In particular, the distance modulus from the recent near-IR PL of \citet{2026AJ....172..148B}, $\mu_0 =18.472\pm0.035$ mag, is in excellent agreement with our value for Reticulum of $18.47\pm0.09$ mag, with our slightly larger uncertainty being dominated by the uncertainty in the zero point of \citet{Neeley2017zero_point} of $\pm0.089$ mag. Figures~\ref{fig:Retic_compare}b and \ref{fig:Retic_compare}c make the same comparison for NGC~1841 and NGC~1466, and while there are fewer historic distance estimates for these two clusters our work shows good agreement with the consensus. 

\subsection{Free slope PL fitting}
\label{sec:free_slope}
\begin{figure}
\centering
\includegraphics[width=\columnwidth]{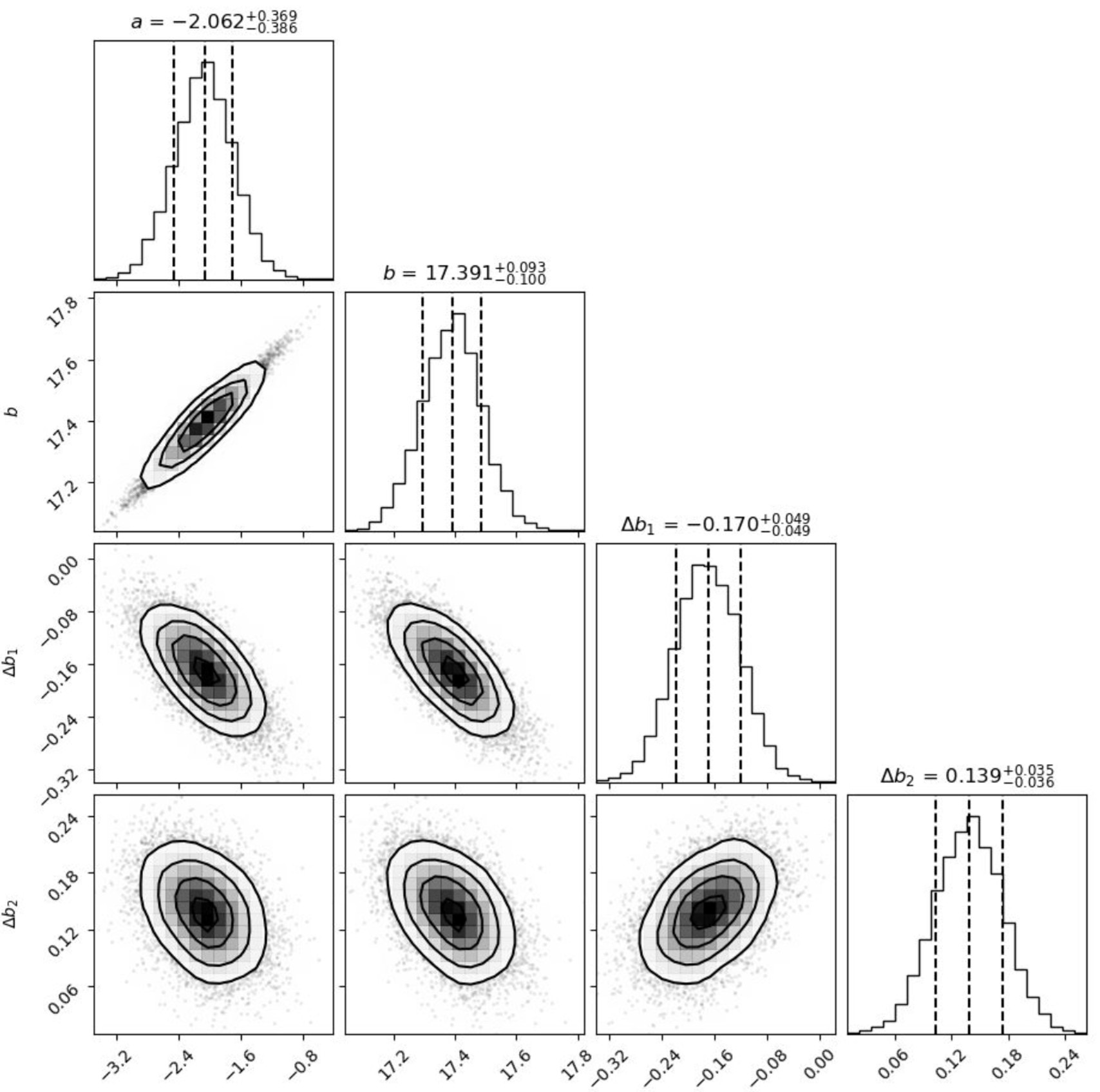}
\caption{Corner plot for the simultaneous three cluster fit PL with slope as a free parameter. $a$ is the slope of the PL, $b$ is intercept of the PL, while $\Delta b_1$ and $\Delta b_2$ are the offsets from Reticulum for NGC~1841 and NGC~1466 respectively.}
\label{fig:Three cluster PL free}
\end{figure}

We repeated the Bayesian linear regression model to simultaneously fit the PL relation in Equation~\ref{eq:Three cluster PL equation} with slope as an additional free parameter to provide an LMC-derived PL. Flat priors of $\pm50$ were applied to the intercept $b$, $\pm20$ to the slope, and $\pm10$ to $\Delta b$.  The corner plot for the free slope LMC-derived PL fit is shown in Figure~\ref{fig:Three cluster PL free}. We find slope $a=-2.06\pm0.38$ consistent with the Galactic GC derived slope of \citet{Neeley2017zero_point}, intercept $b=17.39\pm0.10$ consistent with our fixed slope fit, and the offsets to NGC~1841 $\Delta b_1= -0.17\pm0.05$ and NGC~1466 $\Delta b_2= 0.14\pm0.04$ both consistent with our fixed slope fit. Values are reported as the median of the posterior distribution with uncertainties reported as the 16th and 84th percentiles of the distribution.

\section{Period-Luminosity-Metallicity relations}
\label{sec:PLZ relations}
In Section~\ref{sec:Role of met} we discussed the debate in the literature over the role that metallicity plays in the near-IR PL. In order to address this important question we sought to develop our own empirical PLZ for the three clusters. 

There is limited metallicity data available for the old clusters of the LMC. In particular, there is a lack of spectroscopic metallicities for old cluster RRL, hence indirect techniques have often been used. \cite{2014MNRAS.438.2440D} used \textit{I}-band light curve analysis of 17,337 RRab stars from the OGLE-III catalogue to obtain metallicities for individual stars. This was limited to the OGLE dataset which, while covering 40 square degrees, does not provide full coverage of the LMC halo where our clusters reside. \cite{Narloch2022} used Str\"{o}mgren photometry on individual stars in LMC clusters and obtained mean metallicities for 110 clusters, of which only NGC~1841 is in common with our sample. \cite{14clusters} undertook Period-Amplitude-Metallicity (PAZ) and Period-Metallicity (PZ) studies of RR Lyrae stars in the 14 old clusters of the LMC known to be hosts to these stars. 

Here we adopt the method from \citet[][hereafter M25]{Muraveva_2025_metal}, who used machine learning methods on a large sample of RRL in the \textit{Gaia} DR3 dataset and a subset with spectroscopic metallicities to develop an empirical relationship between [Fe/H], period and the $\phi_{31}$ \textit{G}-band Fourier parameter for RRab stars. Their technique performed well when compared to other spectroscopic data sets, and yielded an RMS error of predicted RRab metallicity values in the training data of 0.28 dex, which is of the order expected for uncertainty in low-resolution and intermediate-resolution spectroscopic metallicities. Adopting the \citetalias{Muraveva_2025_metal} method allows us to find metallicity estimates for individual RRab stars in all of our clusters, and can be used to find metallicities of RRab in other clusters and in the Milky Way on a consistent scale in future work. 

For each RRab in our sample, we derive the metallicity using the \textit{Gaia} DR3 $\phi_{31,G}$ and periods from the \textit{Gaia} {\tt{gaiadr3.vari-rrlyrae}} table. In order to compare our results with the \citet{14clusters} mean cluster metallicities, we convert the \citeauthor{14clusters} values from the \citet{1984ApJS...55...45Z} scale to \citet{2009A&A...508..695C} using
\begin{equation}   
    \label{eq:Carretta}
    \mathrm{[Fe / H]}_{\text{C09}} = 1.105 \mathrm{[Fe/H]}_{\text{ZW84}} +0.160
\end{equation}
and then adding 0.08 dex as per \cite{2021ApJ...908...20C} and \citet{2021ApJ...912..144M} to obtain values on the \cite{2021ApJ...908...20C} scale. The comparison is shown in Table~\ref{tab:Metallicity comparison}. The mean cluster [Fe/H] in this work is shown here for all \textit{Gaia} RRab within 3 arcmin of the cluster centre, demonstrating that the \citet{14clusters} values are consistent with the estimates used in this study in all cases apart from the value for Reticulum derived from \citet{1992AJ....103.1166W}, which is still inside the RMS error from \citet{Muraveva_2025_metal} training data.

The mean cluster metallicies we obtain are only used for the comparison with \citet{14clusters} as shown in Table~\ref{tab:Metallicity comparison}. Inherent degenaracies mean they cannot be used in the PLZ shown in Equation~\ref{eq:Influence of gamma}. $[\mathrm{Fe/H}]_{\textrm{cluster}}$ is constant within each cluster, meaning that $c\times [\mathrm{Fe/H}]_{\textrm{cluster}}$ and the offset between clusters $\Delta b_{\mathrm{cluster}}$ both behave as cluster-level vertical shifts on the PLZ. So with only 3 clusters, $c$ and the cluster offsets will be strongly degenerate. Instead, our PLZ study uses individual RRab metallicities in a methodology similar to that employed by \citet{2026A&A...706A..87L}.

The quality of the \emph{Gaia} light curves is key to our approach as it relies on the $\phi_{31}$ \textit{G}-band Fourier parameter. In Section~\ref{pulsation periods} we discussed our restriction of the RRab sample to those with clean \emph{Gaia} light curves - additionaly a small number of RRab with no $\phi_{31}$ parameter are also dropped for the PLZ analysis (three for Reticulum and one for NGC~1466).

\begin{table*}
	\centering
	\caption{Mean cluster metallicity comparison with \citet{14clusters}. (Reference column indicates the source of photometry used by \citeauthor{14clusters}). Our mean cluster metallicities are from all RRab within 3 arcmin of the cluster centres with $\phi_{31}$ parameter reported in the \textit{Gaia} DR3 {\tt{gaiadr3.vari-rrlyrae}} table.}
	\label{tab:Metallicity comparison}
	\begin{tabular}{lcccccr} 
		\hline
		Cluster & $\mathrm{[Fe/H]}_{\mathrm{C}09}$ & $n_{\text{RRab}}$ \textit{Gaia} & $\mathrm{[Fe/H]}_{\mathrm{C}09}$ & Reference & $n_{\text{RRab}}$ &$\Delta\mathrm{[Fe/H]}_{\mathrm{C}09}$ \\
         & This work & (3 arcmin) & Sarajedini 24 & Sarajedini 24 & Sarajedini 24 & S24$-$This work \\
		\hline
        Reticulum & $-1.60\pm0.10$ & 18& $-1.37\pm0.04$ & \citet{1992AJ....103.1166W}& 20 & $+0.23\pm0.11$ \\
             &  & & $-1.48\pm0.03$ & \citet{2013AJ....145..160K} & 18 &$+0.12\pm0.10$\\
            \hline
            NGC~1841 & $-2.06\pm0.11$ & 18& $-2.08\pm0.06$ & \citet{1990AJ....100.1532W}& 13 & $-0.02\pm0.13$ \\
            \hline
            NGC~1466 & $-1.62\pm0.11$ & 14& $-1.52\pm0.06$ & \citet{1992AJ....104.1395W}& 21 & $+0.10\pm0.13$ \\
            & & &  $-1.64 \pm0.04$ & \citet{2011AJ....142..107K}& 18 & $-0.02\pm0.12$ \\
             & & &  $-1.47\pm0.04$ & \citet{2016AcA....66..131S} & 44& $+0.15\pm0.12$ \\
		\hline
	\end{tabular}
\end{table*}

\begin{figure*}
        \centering
        \includegraphics[width=1.\textwidth]{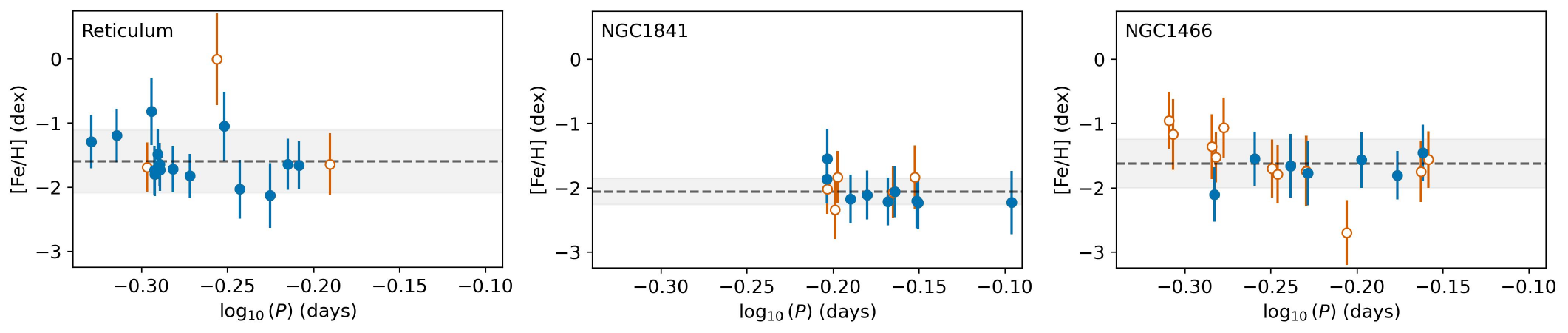}
        \caption{Metallicities for the three clusters found using the empirical formula presented in \citetalias{Muraveva_2025_metal}. Values are [Fe/H] (dex) on the \citet{2021ApJ...908...20C} scale. All RRab within 3 arcmin of the cluster centres with $\phi_{31}$ parameter reported in the \textit{Gaia} DR3 {\tt{gaiadr3.vari-rrlyrae}} table are shown. Blue filled circles indicate RRab for which we recovered mean magnitudes and which were used in the PL, other RRab are shown by orange open circles. The dashed line shows mean value for all data points, and grey band indicates standard deviation as reported in Table~\ref{tab:Metallicity comparison}.}
        \label{fig:Cluster_metallicity}
\end{figure*}

\subsection{Bayesian linear regression analysis}
\label{sec:PLZ fitting}
We again used a Bayesian linear regression model to simultaneously fit the basic PLZ relation:
\begin{equation}   
    \label{eq:plz_bayes}
    m = a \log(P) + b + c \mathrm{[Fe/H]}
\end{equation}
where $c$ is a metallicity coefficient. Similar to the PL fit, we added a parameter to the model to represent the offset for each cluster $(\Delta b)$, and fixed the offset for Reticulum to zero: 
\begin{equation}  \label{eq:Three cluster PLZ equation}
m_{[3.6]ij} = a\log(P) + (b + \Delta b_{j}) + c \mathrm{[Fe/H]}_{ij},
\end{equation}
where the indices $_{ij}$ denote the $i$\textsuperscript{th} star in the $j$\textsuperscript{th} cluster.

\begin{figure}
        \centering
        \includegraphics[width=1.\columnwidth]{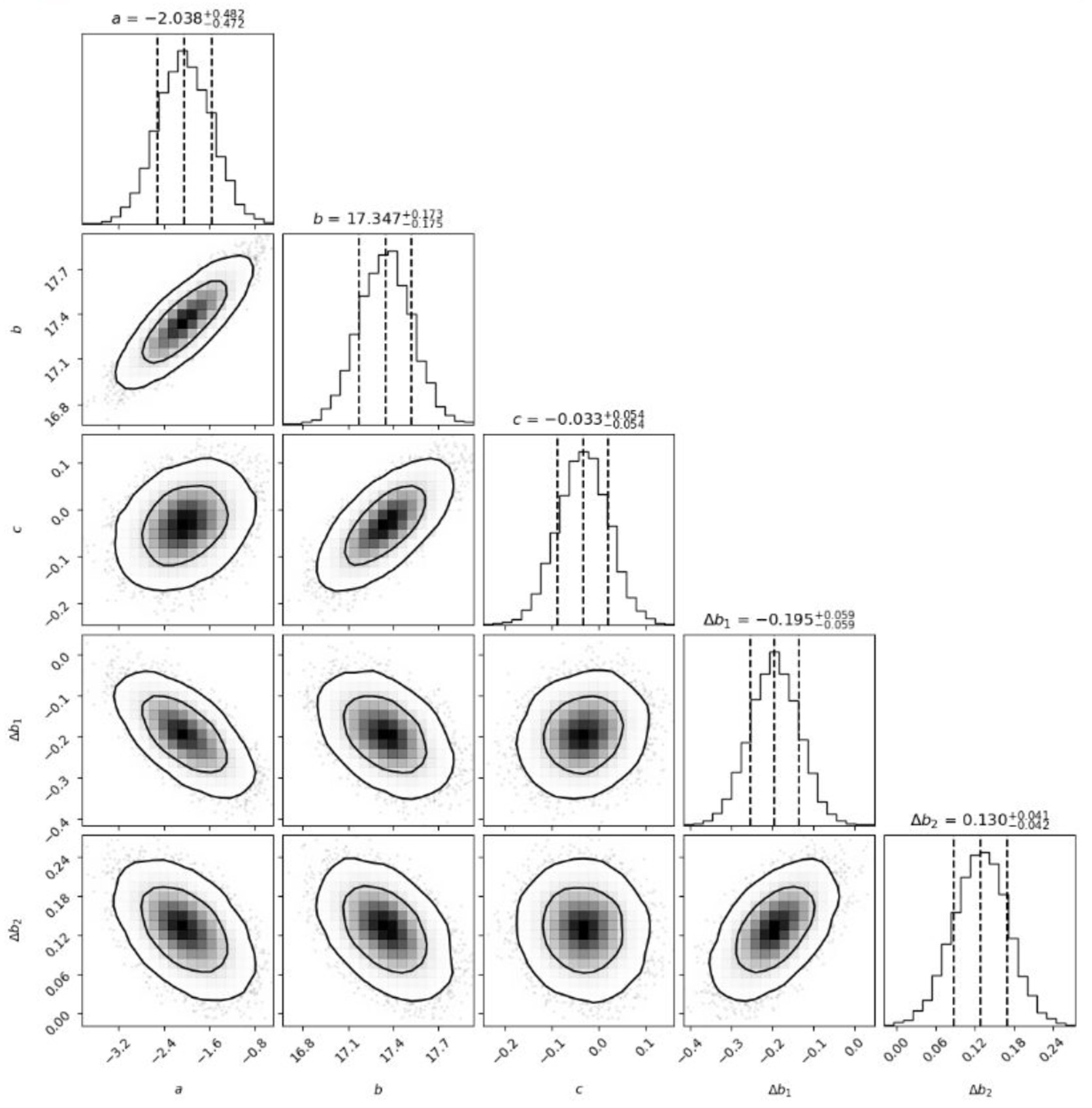}
        \caption{Corner plot for the PLZ showing representations of the posterior distributions of the parameters $a$ (slope), $b$ (intercept), $c$ (metallicity coefficient), $\Delta b_1$ (offset for NGC~1841), $\Delta b_2$ (offset for NGC~1466), and their cross-correlations.}
        \label{fig:three_cluster_plz}
\end{figure}

We used flat priors of $\pm50$ on the independent variables $a$, $b$ and $c$, and $\pm10$ on $\Delta b$, to again allow the full parameter space to be explored. MCMC sampling was used to derive the posterior distribution for these coefficients. The resulting corner plot is provided in Figure~\ref{fig:three_cluster_plz}. We find slope $a=-2.04\pm0.48$ consistent with the Galactic-GC derived slope of \citet{Neeley2017zero_point}, intercept $b=17.35\pm0.17$, and the offsets to NGC~1841 $\Delta b_1= -0.20\pm0.06$ and NGC~1466 $\Delta b_2= 0.13\pm0.04$. We find the metallicity coefficient $c= -0.033\pm0.054$, consistent with zero. Values are reported as the median of the posterior distribution with uncertainties reported as the 16th and 84th percentiles of the distribution. Our finding of negligible dependence of the PL on metallicity ($|c| <0.1$ mag dex$^{-1})$ is in line with several empirical studies cited earlier \citep[e.g][]{2008MNRAS.384.1583S, 2009A&A...502..505B, 2015ApJ...807..127M}.

\section{The intrinsic width of the RRL Period-Luminosity relation}
\label{sec:IW}
The analysis presented earlier in this section follows the standard method for fitting RRL PL relations, i.e. assuming that any dispersion around the mean PL is due to photometric uncertainties. While photometric uncertainties do contribute to the observed PL width, there may also be increased dispersion due to factors inherent to the sample (e.g. line-of-sight depth of the cluster or varying extinction), or distributions in the stellar parameters (e.g. spreads in metallicity, age, mass or temperature). 

The observed width of the RRL PL determines the precision with which the intercept can be measured, which in turn places a limit on the precision of PL-derived distances. In order to determine if the intrinsic width of our PL and PLZ are significant, we repeated the Bayesian linear regression analysis with models that assume the intrinsic (i.e.~error-free) data follows a Gaussian distribution with variance, $V$, encoding scatter in mean luminosity at a given period. The observed PL width is then a result of convolving this Gaussian with the photometric uncertainties. The intrinsic width, $W$, is therefore simply $\sqrt{V}$. Our models treat each data point as being drawn from this projected distribution function by including including $W$ in the inverse variance $(w)$,
\begin{equation}   
    \label{eq:ivar}
    w = 1/ (\sigma^2 + W^2) 
\end{equation}
within the likelihood function of our Bayesian linear regression analysis as suggested by \cite{2010arXiv1008.4686H}.\footnote{See also the Python notebook on fitting a line to a function with an intrinsic width by Adrian Price-Whelan \url{https://adrian.pw/blog/fitting-a-line/}}

\subsection{Fits with an intrinsic width}
\begin{figure}
        \centering
        \includegraphics[width=1.\columnwidth]{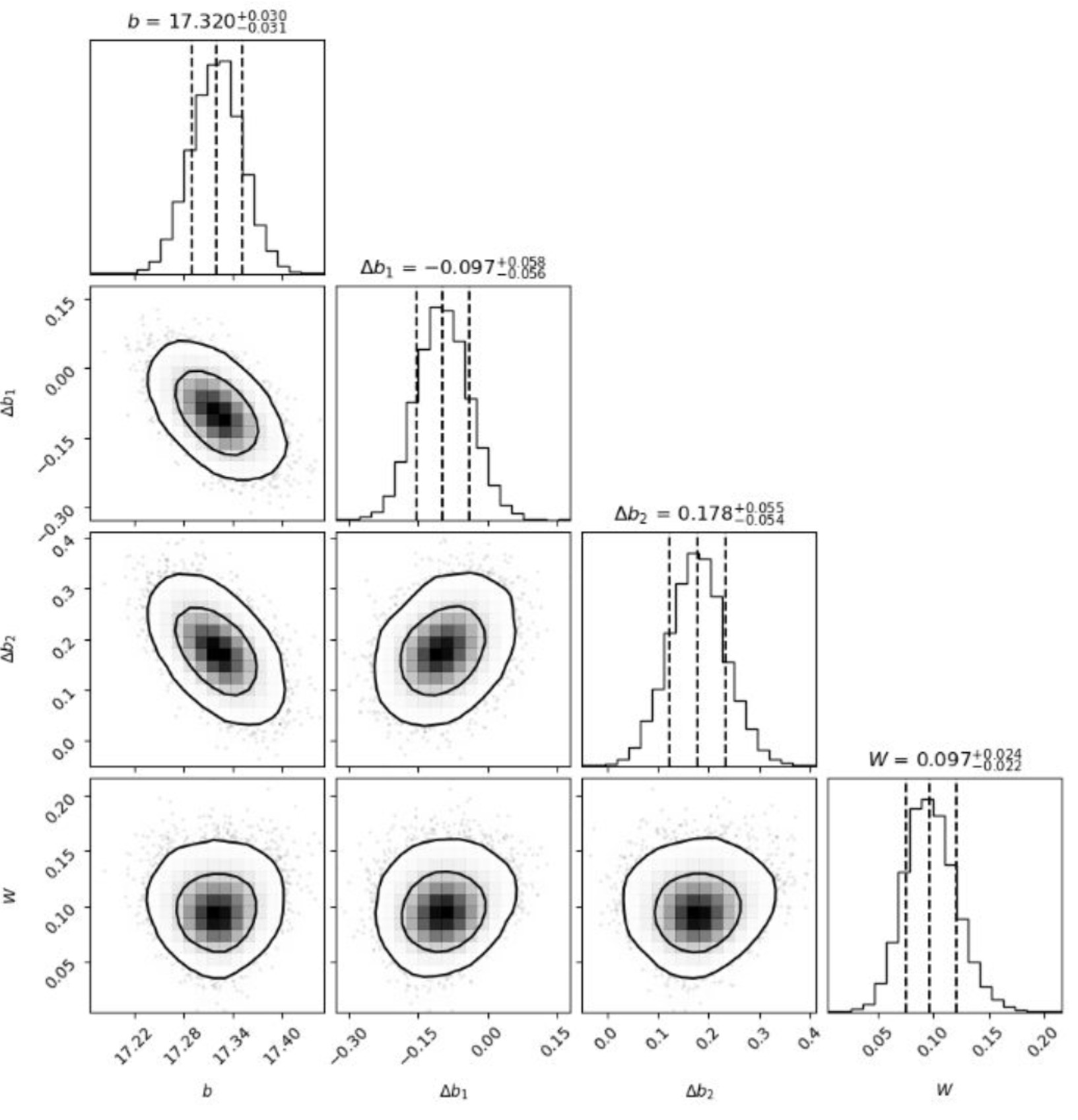}
        \caption{Corner plot for the simultaneous three cluster fit PL with fixed slope, assuming an intrinsic width, showing representations of the posterior distributions of the parameters $b$ (intercept), $\Delta b_1$ (offset for NGC~1841), $\Delta b_2$ (offset for NGC~1466), $W$ (intrinsic width) and their cross-correlations.}
        \label{fig:three_cluster_PL_IW}
\end{figure}

\begin{figure}
        \centering
        \includegraphics[width=1.\columnwidth]{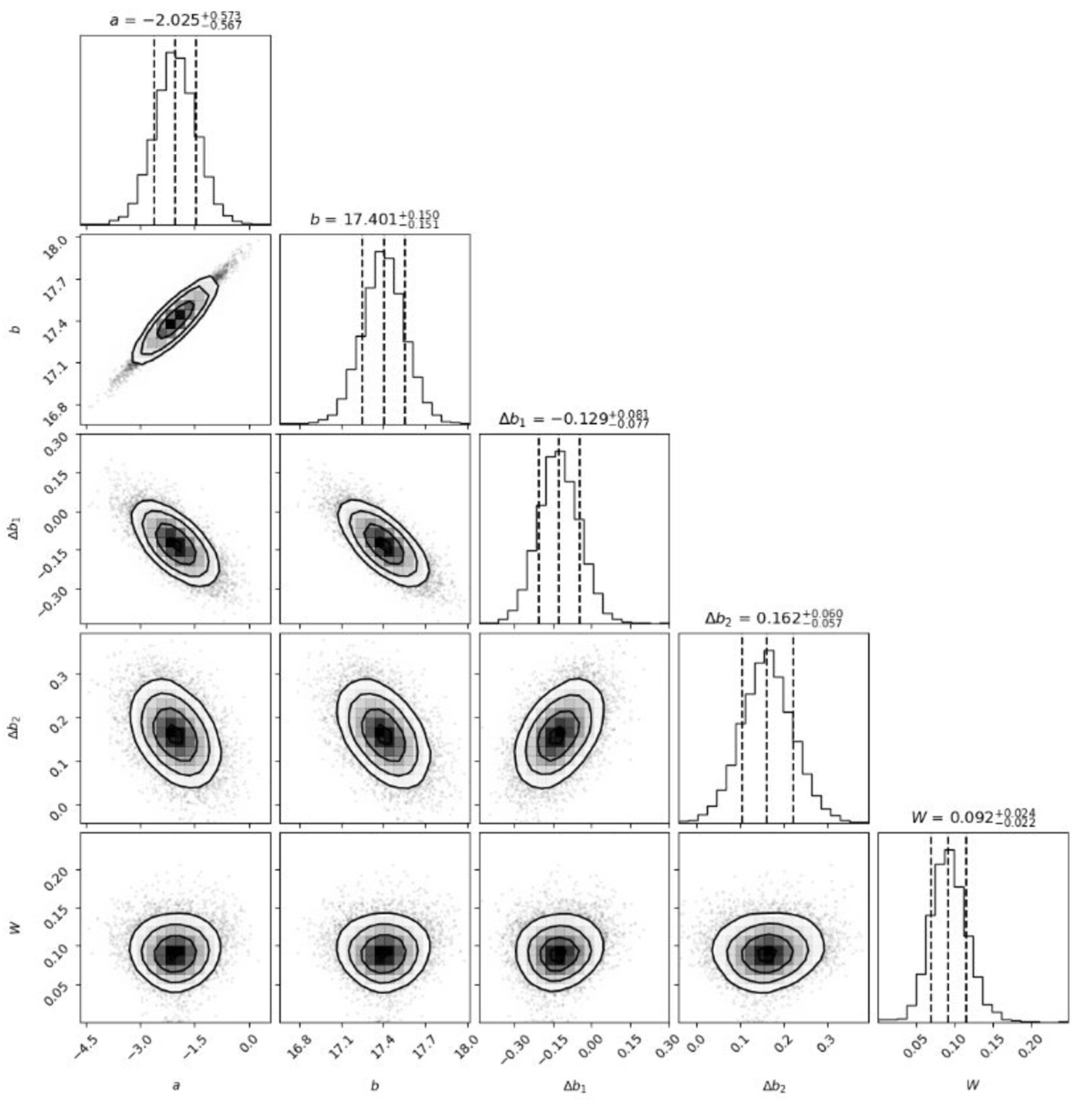}
        \caption{Corner plot for the simultaneous three cluster fit PL with slope as a free parameter, assuming an intrinsic width, showing representations of the posterior distributions of the parameters $a$ (slope), $b$ (intercept), $\Delta b_1$ (offset for NGC~1841), $\Delta b_2$ (offset for NGC~1466), $W$ (intrinsic width) and their cross-correlations.}
        \label{fig:three_cluster_free_PL_IW}
\end{figure}

\begin{figure}
        \centering
        \includegraphics[width=1.\columnwidth]{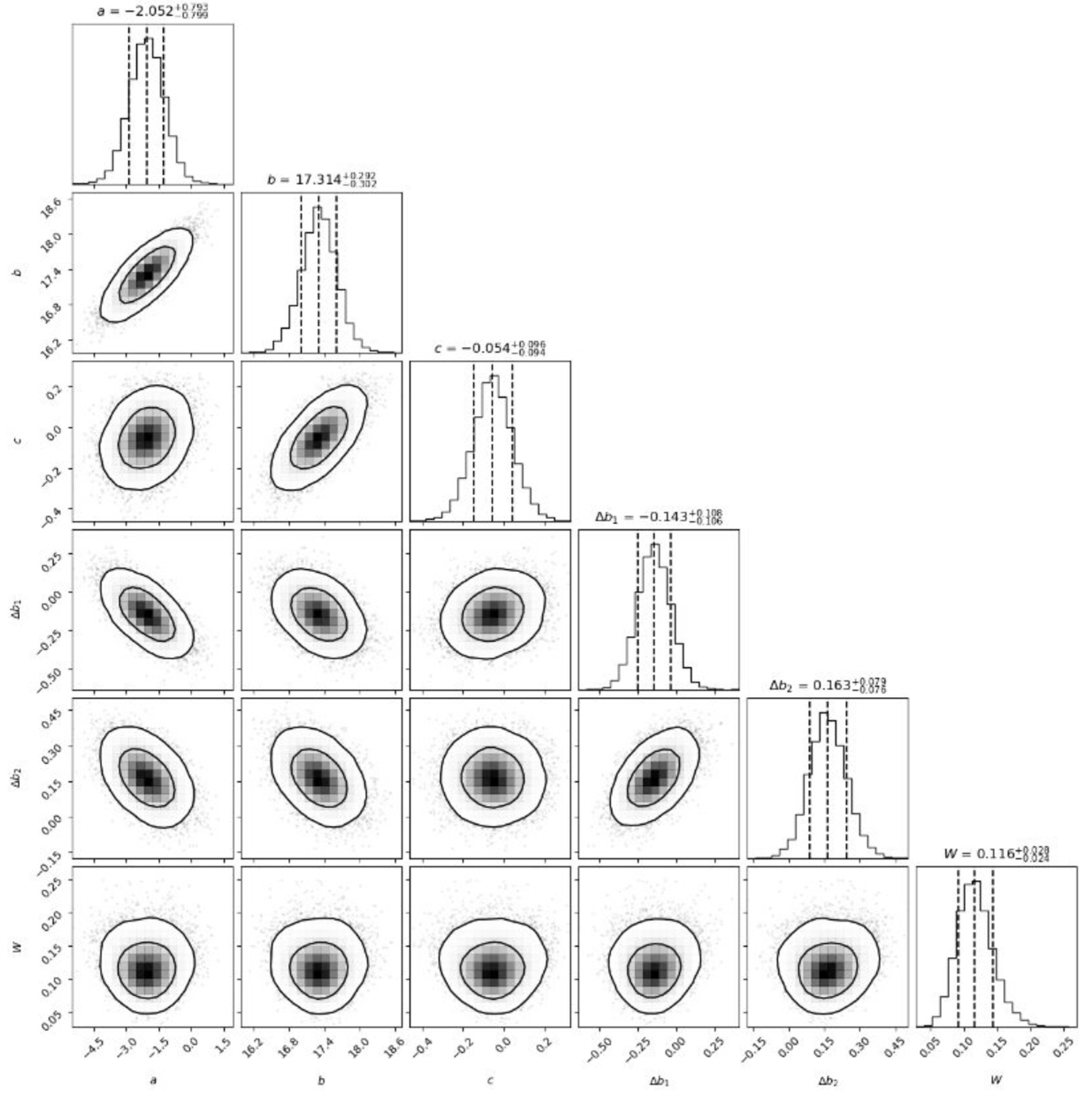}
        \caption{Corner plot for the simultaneous three cluster fit PLZ, assuming an intrinsic width, showing representations of the posterior distributions of the parameters $a$ (slope), $b$ (intercept), $c$ (metallicity coefficient), $\Delta b_1$ (offset for NGC~1841), $\Delta b_2$ (offset for NGC~1466), $W$ (intrinsic width) and their cross-correlations.}
        \label{fig:three_cluster_PLZ_IW}
\end{figure}

We repeated the fits in Sections~\ref{sec:PL Relations} and~\ref{sec:PLZ relations} using the same flat priors, with the addition of intrinsic width as a free parameter with an asymmetric prior to constrain $W$ to non-negative values $(0<W<1)$. Corner plots for the fits with intrinsic width are shown in Figure~\ref{fig:three_cluster_PL_IW} (PL fixed slope), Figure~\ref{fig:three_cluster_free_PL_IW} (PL free slope) and Figure~\ref{fig:three_cluster_PLZ_IW} (PLZ).

We find slopes, intercepts and the offsets to NGC~1841 and NGC~1466 consistent with our fits ignoring IW. For the PLZ we find a metallicity coefficient consistent with our fit ignoring IW. Across all cases we find an intrinsic width $W\approx0.1\pm0.02$~mag.
Our results from the fits incorporating an intrinsic width are shown in Table~\ref{tab:IW results}, with 
values are reported as the median of the posterior distribution with uncertainties reported as the 16th and 84th percentiles of the distribution.

\begin{table*}
	\centering
	\caption{Intrinsic width results}
	\label{tab:IW results}
	{\begin{tabular}{lcccccr} 
		\hline
		Case & Slope & Intercept & $\mathrm{Offset}_{1841}$ & $\mathrm{Offset}_{1466}$ & [Fe/H] coeff & Intrinsic width \\
             & $a$   &   $b$     & $\Delta b_1 $            & $\Delta b_2 $            & $c$          &  $W$\\
        \hline
        PL fixed slope & n/a &  $17.32\pm0.03$ & $-0.10\pm0.06$  & $0.18\pm0.05$    & n/a & $0.10\pm0.02$  \\
        PL free slope   & $-2.03\pm0.57$  & $17.40\pm0.15$ & $-0.13\pm0.08$ & $0.16\pm0.06$ & n/a & $0.10\pm0.02$\\
        PLZ free slope & $-2.05\pm0.79$  & $17.31\pm0.30$ & $-0.14\pm0.11$ & $0.16\pm0.08$ & $-0.054\pm0.095$ & $0.12\pm0.02$ \\
        \end{tabular}
        }
\end{table*}

\section{Conclusions}
\label{sec:Conc}
We present time-series photometry of RRab stars in three old clusters in the LMC, and their flux-weighted mean [3.6] magnitudes derived using \textsc{gloess}. We derived PL and PLZ relations by simultaneously fitting the three clusters using MCMC sampling. For the PL we considered two cases: (i) a PL relation with slope fixed to the value derived for Galactic globular clusters by \citet{Neeley2017zero_point}; and (ii) a PL relation with  slope treated as a free parameter. For the PLZ we treated slope as a free parameter. We repeated these three fits with the addition of an intrinsic width to the relations.

Our photometric sample consists of RRab stars from old clusters in the LMC halo -- a choice made in order to avoid photometric crowding prevalent  in clusters in the LMC's central bar. For this reason, only one LMC cluster (Reticulum) has previously had a mid-IR RRL PL derived. This work, examining three of the old clusters, significantly expands the study of the mid-IR PL and PLZ for the LMC. While the sample size is small in absolute terms, it represents a significant proportion of the RRab stars in the \emph{Gaia} DR3 catalogue for both Reticulum and NGC~1841.

\subsection{Period Luminosity relation and distance modulus}
From our PL with slope fixed to the Galactic-GC derived value in \cite{Neeley2017zero_point} we find distance moduli to the three clusters that are in good agreement with values from other techniques, validating the use of the higher-accuracy Galactic slope in the LMC, and the use of \textsc{photutils} to undertake high-precision time-series photometry in crowded fields. For Reticulum we find $\mu_0=18.47\pm0.09$ mag, for NGC~1841 we find $\mu_0=18.29\pm0.09$ mag, and for NGC~1466 we find $\mu_0=18.61\pm0.09$ mag. Our distance moduli for NGC~1841 and NGC~1466 are the first to be derived from PL relations.

\subsection{The role of metallicity in the PL}
For our simultaneous three-cluster fit, we find the metallicity coefficient $c= -0.033\pm0.054$. This is consistent with zero and can be considered a negligible dependence of the PL on metallicity ($|c| <0.1$ mag dex$^{-1}$).

\subsection{Intrinsic width of the PL and PLZ}
Theory predicts an intrinsic width to the PL of 0.13~mag driven by the spread of metallicities expected for RR Lyrae \citep{Neeley}. In our PLs with Galactic and LMC derived slopes, and in our PLZ, we find an intrinsic width $W\approx0.1\pm0.02$~mag. The fact that we also find no evidence for a metallicity effect in the PL suggests that the dispersion is driven by another factor(s) such as the inherent width of the Instability Strip. Our study differs from previous work in that we restrict the analysis to RRab stars, thereby avoiding the systematic uncertainties associated with fundamentalizing RRc periods. Consequently, our detection of an intrinsic width to the PL cannot be attributed to the systematics introduced by the fundamentalization process.

It should be noted that restricting our selection to RRab stars does not significantly reduce the sample size, with only NGC~1466 hosting RRd, while Reticulum and NGC~1841 host only five and three RRc respectively within the three arcmin field of view.

\subsection{Future work}
Future work will combine these results with complementary mid-IR Cepheid studies such as \cite{2021MNRAS.500..817C} to further unravel the merger history of the Magellanic Clouds. The 1001MC Survey \citep{2019Msngr.175...54C} of the LMC/SMC system to be conducted by the European Southern Observatory's (ESO) 4-metre Multi-Object Spectroscopic Telescope (4MOST) \citep{2019NatAs...3..574D} is anticipated to provide spectroscopic metallicities for tens of thousands of RR Lyrae in these galaxies for the first time. This dramatic increase in the sample size of RR Lyrae with spectroscopic metallicities will enable a more robust spectroscopic calibration of the PLZ, and in the short-term, with targets of opportunity from early operations, will enable an in-depth analysis of the \citetalias{Muraveva_2025_metal} empirical metallicity formula.

\section*{Acknowledgements}
We thank our two anonymous referees for their helpful suggestions.

This work was supported by the Science and Technology Facilities Council [ST/S000526/1] and the lead author is supported by a Science and Technology Facilities Council studentship.

This work has made use of data gathered by the Carnegie RR Lyrae Program (CRRP, PI Freedman) using observations made with the \textit{Spitzer Space Telescope}, which was operated by the Jet Propulsion Laboratory, California Institute of Technology under a contract with NASA. 

This work has made use of data from the European Space Agency (ESA) mission
\textit{Gaia} (\url{https://www.cosmos.esa.int/gaia}), processed by the \textit{Gaia}
Data Processing and Analysis Consortium (DPAC,
\url{https://www.cosmos.esa.int/web/gaia/dpac/consortium}). Funding for the DPAC
has been provided by national institutions, in particular the institutions
participating in the \textit{Gaia} Multilateral Agreement.

This research made use of \textsc{astropy}, a community-developed core \textsc{python} package for Astronomy \citep{astropy}. This research made use of \textsc{photutils}, an \textsc{astropy} package for detection and photometry of astronomical sources \citep{photutils}. This research made use of \textsc{emcee}, a \textsc{python} Markov chain Monte Carlo (MCMC) ensemble sampler \citep{2013PASP..125..306F}.

We thank Simon Jeffery for permission to use an adapted version of the pulsation HR diagram from \citet{2016MNRAS.458.1352J} as Figure~\ref{fig:HR_Diagram} in this work, and for providing an updated version of his original plot.

Our plots follow the Colour Universal Design (CUD) system \citep{article} to ensure they are accessible to all vision types.

\section*{Data Availability}
The \emph{Spitzer} images used in this work were taken as part of the Carnegie RR Lyrae Program (CRRP, PI Freedman, program ID 90088) and are available through the NASA IRSA \emph{Spitzer} Heritage Archive. 

All Warm \emph{Spitzer} calibration images used in this work are available in the Warm \emph{Spitzer} Calibration and Analysis Files section of the \emph{Spitzer} documentation (\url{https://irsa.ipac.caltech.edu/data/SPITZER/docs/irac/calibrationfiles/}).

RR~Lyrae parameters used in this work are from \emph{Gaia} Data Release 3 \texttt{gaia\_source} and \texttt{vari\_rrlyrae} tables, with objects selected as described in Section~\ref{sec:GVS}.



\bibliographystyle{mnras}
\bibliography{example} 




\clearpage
\appendix
\section{Full PL light curve sample}
\label{APP_A}

\begin{figure*}
        \centering
        \includegraphics[width=\textwidth]{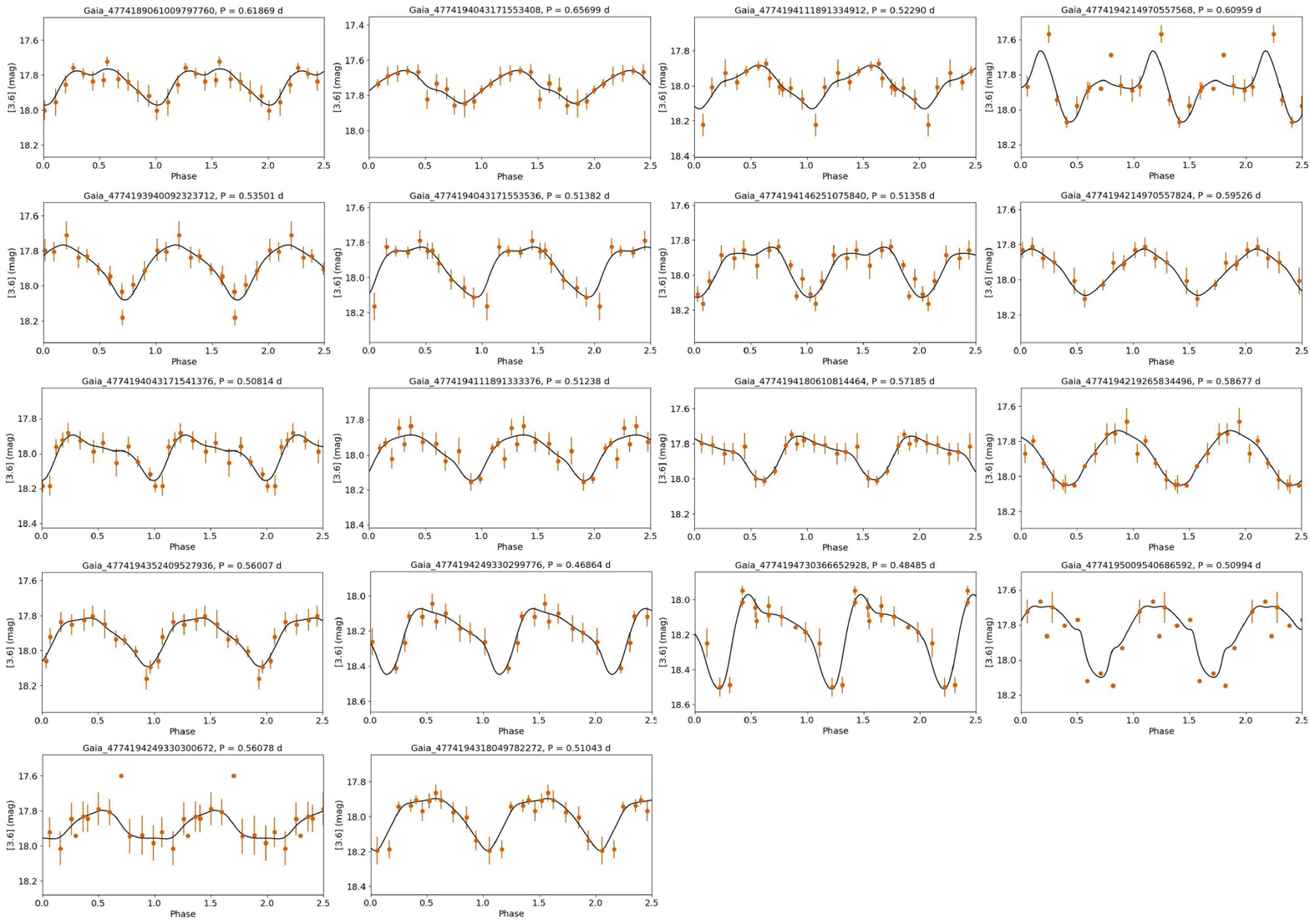}
        \caption{\textsc{gloess} fitted phase-folded light curves for all RRab in the Reticulum PL. The orange data points and errorbars are our photometric results for each epoch, the black light curve is the fit provided by GLOESS to this dataset. A common magnitude range is used on all light curves. Note that some photometric data points do not have errorbars as the EPSF fitter was unable to provide them, however we have included these points for completeness.}
        \label{fig:Retic_all_LC}
\end{figure*}

\begin{figure*}
        \centering
        \includegraphics[width=\textwidth]{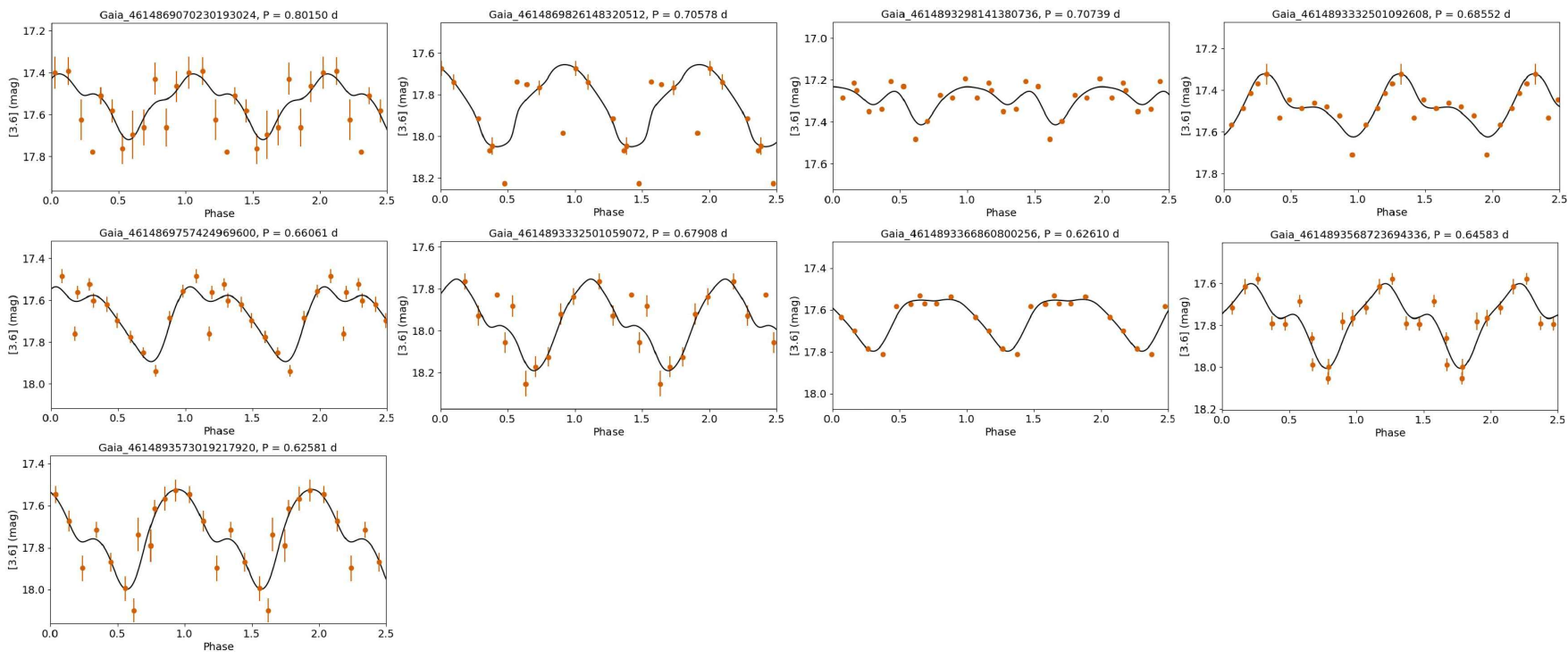}
        \caption{\textsc{gloess} fitted phase-folded light curves for all RRab in the  NGC~1841 PL. The orange data points and errorbars are our photometric results for each epoch, the black light curve is the fit provided by GLOESS to this dataset. A common magnitude range is used on all light curves. Note that some photometric data points do not have errorbars as the EPSF fitter was unable to provide them, however we have included these points for completeness.}
        \label{fig:1841_all_LC}
\end{figure*}

\begin{figure*}
        \centering
        \includegraphics[width=\textwidth]{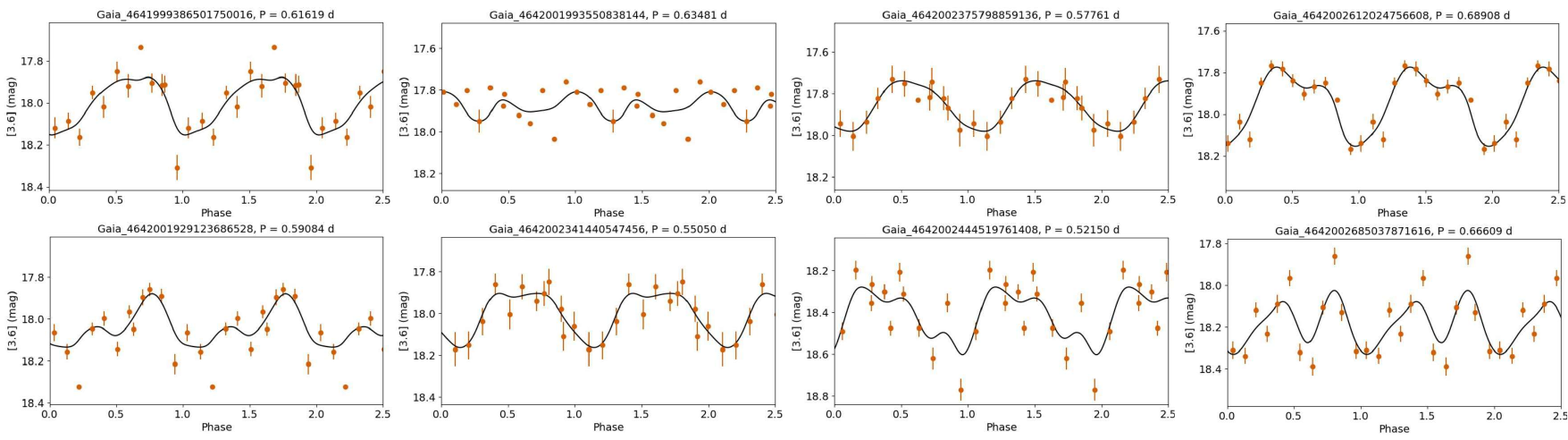}
        \caption{\textsc{gloess} fitted phase-folded light curves for all RRab in the NGC~1466 PL. The orange data points and errorbars are our photometric results for each epoch, the black light curve is the fit provided by GLOESS to this dataset. A common magnitude range is used on all light curves. Note that some photometric data points do not have errorbars as the EPSF fitter was unable to provide them, however we have included these points for completeness.}
        \label{fig:1466_all_LC}
\end{figure*}

Light curves for the full sample of RRab used in the construction of the PL are presented in Figures~\ref{fig:Retic_all_LC} (Reticulum), \ref{fig:1841_all_LC} (NGC~1841) and \ref{fig:1466_all_LC} (NGC~1466). Flux-weighted mean magnitudes and uncertainties derived from the \textsc{gloess} light curves are given in Table~\ref{tab:mean_mags}.

\begin{table}
	\caption{Flux-weighted mean magnitudes and uncertainties derived from \textsc{gloess} light curves. Right-most column shows how these recovered RRL were used in this work, PL and PLZ, PL only, or unused.}
	\label{tab:mean_mags}
	\begin{tabular}{lrrrrr} 
		\hline
		Cluster& \emph{Gaia}~DR3 Source ID & \emph{Gaia}~DR3 Period & [3.6] & $\sigma_{[3.6]}$ & Use\\
            &  & (days) & (mag) & (mag)  \\
		\hline
Reticulum & 4774189061009797760 & 0.61869 & 17.841 & 0.062 & PL/PLZ\\ 
 & 4774193940092323712 & 0.53501 & 17.891 & 0.096 & PL/PLZ\\ 
 & 4774194043171541376 & 0.50814 & 17.998 & 0.076 & PL/PLZ\\ 
 & 4774194043171553408 & 0.65699 & 17.748 & 0.054 & PL\\ 
 & 4774194043171553536 & 0.51382 & 17.937 & 0.085 & PL/PLZ\\ 
 & 4774194111891333376 & 0.51238 & 17.984 & 0.076 & PL/PLZ\\ 
 & 4774194111891334912 & 0.52290 & 17.991 & 0.075 & PL/PLZ\\ 
 & 4774194146251075840 & 0.51358 & 17.947 & 0.083 & PL/PLZ\\ 
 & 4774194180610814464 & 0.57185 & 17.859 & 0.072 & PL/PLZ\\ 
 & 4774194214970557568 & 0.60959 & 17.867 & 0.119 & PL/PLZ\\ 
 & 4774194214970557824 & 0.59526 & 17.945 & 0.077 & PL/PLZ\\ 
 & 4774194219265834496 & 0.58677 & 17.877 & 0.091 & PL \\ 
 & 4774194249330299776 & 0.46864 & 18.206 & 0.108 & PL/PLZ\\ 
 & 4774194249330300672 & 0.56078 & 17.891 & 0.070 & PL\\ 
 & 4774194318049782272 & 0.51043 & 17.999 & 0.086 & PL/PLZ\\ 
 & 4774194352409527936 & 0.56007 & 17.917 & 0.082 & PL/PLZ\\ 
 & 4774194730366652288 & 0.61902 & 17.925 & 0.092 & --\\ 
 & 4774194730366652928 & 0.48485 & 18.173 & 0.156 & PL/PLZ\\ 
 & 4774195009540686592 & 0.50994 & 17.837 & 0.118 & PL/PLZ\\ 
 & 4774195073964030592 & 0.55458 & 17.911 & 0.039 & --\\ 
		\hline
NGC~1841 & 4614869070230193024 & 0.80150 & 17.538 & 0.092 & PL/PLZ\\ 
 & 4614869757424969600 & 0.66061 & 17.677 & 0.105 & PL/PLZ\\ 
 & 4614869826148320512 & 0.70578 & 17.823 & 0.118 & PL/PLZ\\ 
 & 4614869899160800512 & 0.68391 & 17.403 & 0.109 & --\\ 
 & 4614893298141380736 & 0.70739 & 17.294 & 0.057 & PL/PLZ\\ 
 & 4614893332501059072 & 0.67908 & 17.955 & 0.127 & PL/PLZ\\ 
 & 4614893332501092608 & 0.68552 & 17.477 & 0.089 & PL/PLZ\\ 
 & 4614893366860800256 & 0.62610 & 17.630 & 0.072 & PL/PLZ\\ 
 & 4614893401221834240 & 0.63255 & 17.700 & 0.129 & --\\ 
 & 4614893568723694336 & 0.64583 & 17.774 & 0.118 & PL/PLZ\\ 
 & 4614893573019217920 & 0.62581 & 17.726 & 0.139 & PL/PLZ\\ 
		\hline
NGC~1466 & 4641999386501750016 & 0.61619 & 17.990 & 0.100 & PL\\ 
 & 4642001929123686528 & 0.59084 & 18.033 & 0.085 & PL/PLZ\\ 
 & 4642001993550838144 & 0.63481 & 17.879 & 0.049 & PL/PLZ\\ 
 & 4642002341440547456 & 0.55050 & 17.995 & 0.076 & PL/PLZ\\ 
 & 4642002375798859136 & 0.57761 & 17.850 & 0.070 & PL/PLZ\\ 
 & 4642002444519761408 & 0.52150 & 18.416 & 0.107 & PL/PLZ\\ 
 & 4642002577662963968 & 0.57239 & 18.196 & 0.095 & --\\ 
 & 4642002612024756608 & 0.68908 & 17.936 & 0.110 & PL/PLZ\\ 
 & 4642002685037871616 & 0.66609 & 18.179 & 0.099 & PL/PLZ\\ 
        \hline
	\end{tabular}
\end{table}

\bsp	
\label{lastpage}
\end{document}